\documentclass[10pt,journal,compsoc]{IEEEtran}

\usepackage{setspace}
\usepackage{tabularx}
\usepackage{amsmath}
\usepackage{amssymb}
\usepackage{subfigure}
\usepackage[dvips]{graphics} 
\usepackage{epstopdf}
\usepackage{epsfig}
\usepackage{float}
\usepackage{array,longtable}
\usepackage{algorithm}
\usepackage{algorithmic}
\usepackage[utf8]{inputenc}
\usepackage{textcomp}
\usepackage{array}
\usepackage{url}
\usepackage{epsfig}
\usepackage{pstricks}
\usepackage{graphicx}

\begin{document}

\title{Discrete-Time Ping-pong Optimized Pulse Shaping-OFDM (POPS-OFDM) Operating on Time and Frequency Dispersive Channels \\ for 5G Systems}


\author{\IEEEauthorblockN{Zeineb Hraiech, Mohamed Siala and Fatma Abdelkefi \\}
\IEEEauthorblockA{ MEDIATRON Laboratory, SUP’COM, University of Carthage, Tunisia\\ Email: \{zeineb.hraiech, mohamed.siala, fatma.abdelkefi\}@supcom.tn
}
}

\IEEEtitleabstractindextext{%
\begin{abstract}
The Fourth Generation (4G) of mobile communication systems was optimized to offer high data rates with high terminal mobility by ensuring strict synchronism and perfect orthogonality. However, the trend for novel applications, that had not been feasible a few years back, reveals major limits of this strict synchronism and imposes new challenges and severe requirements. Among those, the sporadic access generated by Machine-Type Communication (MTC) and the exchange of small data packets with small payloads, respecting perfect synchronization procedure, imposes the use of large overhead compared to the useful data and hence lead to a significant system performance degradation. As a consequence, MTC nodes need to be coarsely synchronized to reduce the signaling load while reducing the round-trip delay. However, coarse synchronization can dramatically damage the waveforms orthogonality in the Orthogonal Frequency Division Multiplexing (OFDM) signals, which results in oppressive Inter-Carrier Interference (ICI) and Inter-Symbol Interference (ISI). Moreover, in 4G, tremendous efforts must be spent to enhance system performance under strict synchronism in collaborative schemes, such as Coordinated Multi-Point (CoMP). As a consequence, the use of non-orthogonal waveforms becomes further essential in order to meet the upcoming requirements.
In this context, we propose here a novel waveform construction, referred to Ping-pong Optimized Pulse Shaping-OFDM (POPS-OFDM), which is believed to be an attractive candidate for the optimization of the radio interface of next 5G mobile communication systems. Through a maximization of the Signal to Interference plus Noise Ratio (SINR), this approach allows optimal and straightforward waveform design for multicarrier systems at the Transmitter (Tx) and Receiver (Rx) sides. Furthermore, the optimized waveforms at both Tx/Rx sides have the advantage to be adapted to the channel propagation conditions and the impairments caused by strict synchronism relaxation for reduced synchronization overhead. Hence, this straightforward optimization results in spectacular performance enhancement compared to classical multicarrier systems using conventional waveforms. In this paper, we analyze several characteristics of the proposed waveforms and shed light on relevant features, which make it a powerful candidate for the design of 5G system radio interface waveforms.
\end{abstract}
}

\maketitle

\IEEEdisplaynontitleabstractindextext

\IEEEpeerreviewmaketitle


\IEEEraisesectionheading{\section{Introduction}\label{sec:introduction}}
\IEEEPARstart{R}{ecently}, a potential research confirms that transition to the next Fifth Generation (5G) of mobile communication systems becomes further essential in order to meet the future communication services and needs. First and foremost, the trend to offer novel applications of wireless cellular systems that are expected to be feasible by 2020 certainly plays a key role in future communications drivers and imposes new challenges. Among those applications, the Tactile Internet [1-3], which comprises real-time applications with extremely low latency requirements, imposes a time budget on the physical layer below 100 $\mu s$ [3]. As a consequence, the exchange of small data packets with small payloads, respecting a perfect synchronization procedure, imposes an intensive exchange of signaling messages such as synchronization and channel estimation pilot sequences. Hence, it incurs a large latency which leads to a significant system performance degradation in terms of efficient use of energy and radio resources. Coarse synchronization can solve the problem and reduce the signaling load while reducing the round-trip delay. 
In addition to the latter, the Internet of Things (IoT) \cite{ref6}, which generates sporadic access, constitutes another significant challenge research domain [1-4]. Besides a scalability problem, MTC devices need to be coarsely synchronized to guarantee long lifetimes. However, strict synchronism relaxation can dramatically damage the waveforms orthogonality in the Orthogonal Frequency Division Multiplexing (OFDM) signals, on which many recent wireless standards rely \cite{zheng}, since it results in oppressive Inter-Carrier Interference (ICI) and Inter-Symbol Interference (ISI). In addition to that, in 4G, several transmission schemes and applications require strict obedience to synchronism procedure. For example, evolved Nodes B (eNBs) transmitting Multimedia Broadcast over Single Frequency Network (MBSFN), should guarantee perfect synchronism to achieve high operating SNR. Hence, in order to compensate the differences of the propagation delays between eNBs transmitting MBFSN, the adjunction of extended Cyclic Prefix (CP), which is greater than the CP used by default for the channel equalization in the frequency domain, further constrain the bandwidth efficiency \cite{ref10,ref101} and reduce data capacity. Also, tremendous efforts must be spent to enhance system performance under strict synchronism in collaborative schemes, such as Coordinated Multi Point (CoMP).
\newline
In addition to its sensitivity to synchronism and waveform orthogonality, several research studies highlighted a set of drawbacks of OFDM transmission, including mainly the out-of-band emissions, which can result in an important interference level between systems using adjacent bands. Moreover, Carrier Frequency Offset (CFO), which requires sophisticated synchronization mechanisms to guarantee that the orthogonality is not affected \cite{comp}, is one of most well-known disturbances for OFDM. As a consequence, 4G systems can't match the needs cited above, since they require tight synchronization and incur high latencies. Therefore, a non-orthogonal future wireless multi-carrier scheme with well-localized waveforms in time and frequency domains would represent an interesting candidate to be used for the 5G systems. These novel waveforms should be robust to relaxed time/frequency synchronization requirements and imperfect channel state information.
\newline
Besides wireless radio frequency applications stated above, many works shed lights on underwater acoustic applications such as submarine multimedia surveillance, undersea explorations, video-assisted navigation and environmental monitoring \cite{ref8} as an actual and important applications domain. However, the underwater acoustic channels are generally recognized as one of the most difficult communication media in use today \cite{ref8}, \cite{ref9}. The acoustic propagation suffers from severe transmission losses, time-varying multipath propagation, high Doppler spreads, and high propagation delays \cite{ref8}, \cite{ref9}. Thus, in addition to the relaxed synchronism and imperfect orthogonality requirements, this area of applications requires the design of novel waveforms to reduce ISI and ICI and also minimize energy spreading.
\newline
Several solutions were proposed in the literature to mitigate ICI and ISI, when the propagation channel is doubly dispersive and the obedience to strict synchronism and perfect orthogonality are challenged. One of the envisaged solutions is referred to as Generalized Frequency Division Multiplexing (GFDM) \cite{ref10}, \cite{ref11}, which is a new concept for flexible multi-carrier transmission. It offers a low out-of-band radiation of the transmitted signal compared to OFDM, due to an adjustable pulse shaping filter that is applied to the individual subcarriers. 
As such, it strictly avoids harmful interference to legacy TV signals, allowing to opportunistically exploiting spectrum white spaces for wireless data communications. It also features a block-based transmission, using CP insertion, possible windowing for out-of-band (OOB) spectrum leakage, and efficient FFT-based equalization. One of the major drawbacks of GFDM is the requirement to operate on a non-time selective (frequency-dispersive) channel within each transmitted block and to guarantee perfect frequency synchronization at the receiver. Unfortunately, for a given CP insertion overhead, the aggregation of a significant number of transmitted symbols to increase spectrum efficiency makes these assumptions unrealistic or difficult to meet. Another major drawback is due to the sampled nature of GFDM and the use of all potential transmit subcarriers to preserve spectrum efficiency, which breaks waveform spectrum shaping because of spectrum folding. One of the proposed ways to reduce OOB power emission is to apply some kind empirical filtering through a multiplication of each transmitted block by a smooth windowing.
\newline
Another serious multiple radio access candidate under consideration for 5G is Filter Bank Multi-Carrier (FBMC) \cite{ref12}, \cite{ref13}. It has many advantages over OFDM, such as having much better control of the out-of-band radiations due to the frequency localized shaping pulses. This scheme discards the concept of CP and relies on a per-subcarrier equalization to combat ISI and hence improves the spectral efficiency \cite{FBMC}. However, the low latency requirements are not guaranteed by FBMC due to the use of long filter lengths \cite{ref10}, in addition to not guaranteed robustness to time synchronization imperfections.
\newline Furthermore, a Universal Filtered Multi-Carrier (UFMC) \cite{ref14}, \cite{ref5} approach was introduced as an alternative to OFDM where a filtering operation is applied to a group of consecutive subcarriers to reduce the OOB emissions. Here too, UFMC does not require the use of CP, which makes UFMC more sensitive to small time misalignment than CP-OFDM \cite{ref14}. Hence, it will not be suitable for applications which need a relaxed time/frequency synchronization requirements.
\newline
In the same context, a dynamic waveform construction referred as Ping-pong Optimized Pulse Shaping-OFDM (POPS-OFDM) \cite{ref61} was introduced as an attractive candidate for the physical layer of 5G systems in order to eliminate the sensitivity to orthogonality and synchronism constraints. This innovative approach banks on a non-orthogonal wireless multi-carrier scheme which allows the design of well-localized waveforms in both time and frequency domains \cite{ref62}. Through an iterative maximization of the Signal to Interference plus Noise Ratio (SINR), POPS-OFDM deduces the optimized waveforms at Tx/Rx sides. In addition to ICI/ISI mitigation, these waveforms deal efficiently with the problem of the small data packets introduced by the IoT and the MTC in future wireless communication systems, since they are robust against time/frequency synchronization errors. In this paper, we analyze several characteristics of the proposed waveforms and shed light on relevant features.\newline
We note that the conference versions of this paper \cite{ref61} and \cite{ref62} contain parts of the results presented herein. However, we believe that the actual version is a substantial extension of the conference versions, where we detail in a precise way the different concepts. We highlight below the main extra contributions beyond the conference versions:
\begin{itemize}
\item We present a discrete-Time POPS-OFDM version in a general context, without any constraint on the choice of the channel propagation. However, to make the simulations tractable, we consider a radio channel where the scattering function has a multipath power profile with an exponential truncated decaying model and classical Doppler spectrum (Section 6).  
\item We evaluate the complexity of POPS-OFDM implementation through a careful presentation of its underlying methodology.
\item We derive an upper bound on the SINR of POPS-OFDM algorithm and the exact SINR expression of the conventional OFDM systems (Sections 5 and 4.4.1 respectively).
\item  We test the performance of POPS-OFDM for different Transmitter/Receiver (Tx/Rx) pulse shape durations and we highlight the inversion properties in time and hence characterize the properties of the optimization solution, once it was unique.
\item We draw a discussion on the correlation noise and the careful choice of the couple of transmitted and received waveforms or their temporal inverses with exchange of Tx/Rx roles, to reduce noise correlation.
\item We precise carefully the treatment done in cases of classical OFDM, the discovery of the duality of CP-OFDM and ZP-OFDM systems and the fact that CP-OFDM has a zero correlation between the noise samples at the receiver, while ZP-OFDM shows a frequency correlation between the subcarriers of one OFDM symbol which tends, in the absence of a good interleaving, to reduce correction capacity of the used error correcting code.
\item We assess POPS-OFDM sensitivity to an estimation error of the channel spread factor and its robustness against the time and frequency synchronization errors.
\end{itemize}

The rest of this paper is organized as follows. In Section III, we present our multicarrier system model and we specify its transmitter and receiver blocks. We also detail the propagation channel models that will be used in this paper. In Section IV, we focus on the derivation of the SINR expression and describe the iterative POPS-OFDM technique for waveform design. In Section V, we illustrate the obtained optimization numerical results and highlight the efficiency of the proposed POPS-OFDM algorithm. Section VI will be dedicated to evaluate numerically the optimization technique in terms of robustness against time and frequency synchronization errors and its sensibility to waveform initializations. Finally, section VII draws conclusions and perspectives of our work.
\section{Notations}
\begin{center}
\begin{tabular}{| p{0.4\linewidth} || p{0.6\linewidth} |}
\hline Notation& Meaning\\
\hline
\hline  $\Re\{z\}=x$ & The real part of the complex number $z=x+iy$ \\
\hline $\mathbb{E}$ & The expectation operator  \\
\hline $\underline{\underline{V}}=[V_{pq}]_{p,q\in \mathbb{Z}}$& The doubly underlined \underline{\underline{V}} refers to matrix \underline{\underline{V}} with $(p,q)^{th}$ entry $V_{pq}$\\
\hline $\underline{\underline{{{\Phi}}}}_{\nu_k}$&Hermitian matrix with $(p,q)^{th}$ entry $ e^{j2\pi \nu_k T_s (q-p)}$\\
\hline$\underline{V}=[V_q]_{q\in \mathbb{Z}}$& Underlined \underline{V} refers to vector \underline{V} with $q^{th}$ entry $V_q$ \\
\hline ${\underline{\underline{I}}}$&Identity matrix, with ones in the main diagonal and zero elsewhere\\
\hline ${{\sigma}_{p}}(\underline{V}) $ & Time shift by $p$ samples of vector $\underline{V}=[V_q]_{q \in \mathbb{Z}}$, i.e, ${{\sigma}_{p}}(\underline{V})=[V_{q-p}]_{q\in \mathbb{Z}}$\\
\hline $\varpi{(\underline{V})}$ & Temporal inversion of vector $\underline{V}$, i.e, $\varpi{(\underline{V})}=[V_{-q}]_{q\in \mathbb{Z}}$ \\
\hline $\delta(i-j)=\left\{\begin{array}{ll}1 & i=j \\ 0 & i\neq j\end{array}\right.$&Kroncker delta function\\
\hline $.^H$ &  Hermitien transpose operator\\
\hline $.^T$ &  Transpose operator\\
\hline  $.^*$ &  Element-wise conjugation operator\\
\hline $\odot$& Component-wise product of two vectors or matrices\\
\hline $\otimes$& Kronecker product of two vectors ($\underline{X}\otimes\underline{V}=[X_q\underline{V}]_{q \in \mathbb{Z}} $)\\
\hline  $<\underline U,\underline V>$ & Hermitian scalar product of $\underline U$ and $\underline V$\\
\hline  $||\underline V||$ & Norm of a vector $\underline{V}$, $||\underline V||=\sqrt{<\underline V,\underline V>}$\\
\hline
\end{tabular}
\end{center}
\section{System Model}
This section provides preliminary concepts and notations related to the considered multicarrier scheme and the channel assumptions and model. Furthermore, we consider their discrete time version in order to simplify the implementation of this technique and hence, the theoretical derivations that will be investigated.
\subsection{OFDM Transmitter and receiver blocks}
We denote by $T$ the OFDM symbol duration and by $F$ the frequency spacing between two adjacent subcarriers.
The transmitted signal is sampled at a sampling rate $R_{s}=\frac{1}{T_{s}}$, where $T_{s}$ is the sampling period. We choose the symbol duration as an integer multiple of the sampling period, i.e. $T=NT_{s}$ , where $N \in \mathbb{N}^*$. We also choose the subcarrier spacing inverse $1/F$ as an integer multiple of the sampling period, i.e. $1/F=QT_{s}$ , where $Q \in \mathbb{N}^*$.\newline The time-frequency lattice density of the studied OFDM system, which is always taken below unity to account for the equivalent effect of guard interval insertion in conventional OFDM, is defined as
\begin{align}
\Delta=\frac{1}{FT} \nonumber
\end{align}
Hence $$
\Delta=QT_{s}\frac{1}{NT_{s}}=\frac{Q}{N}\leq 1,
$$
which means that $Q$ is always smaller than or equal to $N$ and that the time-frequency lattice density is always rational. The difference $(N-Q)T_{s}$ is equivalent to the notion of guard interval in conventional OFDM.
\newline
Because of sampling, the total spanned bandwidth is equal to $\frac{1}{T_s}$ and therefore the number of subcarriers is finite and equal to ${({1}/{T_s})}/{F} = Q$. Therefore, adopting $[0, \frac{1}{T_s})$ as the spanned frequency band, the subcarrier frequencies are given by $mF=m/QT_{s}$, $m=0,1,..,Q-1$. 
\newline
The sampled version of the transmitted signal is represented by the infinite vector
\begin{align}
{\underline{e}}=(....,e_{-2},e_{-1},e_{0},e_{1},.....)^{T}=[e_{q}]_{q\in \mathbb{Z}} \nonumber
\end{align}
where  $e_{q}$ is the transmitted signal sample at time $qT_{s}$.
\newline
The transmitted signal can be written as
\begin{equation}
{\underline{e}}=\sum_{mn} a_{mn} \ \underline \varphi_{mn}
\end{equation}
where $a_{mn}$ is the transmitted symbol at time $nT$ and frequency $mF$, and $$\underline \varphi_{mn}=[\varphi(q-nN)e^{j2\pi\frac{mq}{Q}}]_{q\in \mathbb{Z}}$$ is the vector used for the transmission of symbol $a_{mn}$, which results from a time shift of $nNT_{s}=nT$ and a frequency shift of $mF=m/QT_{s}$ of the transmission prototype vector $\underline\varphi$.
Also, we denote by $D_{\underline \varphi}T$ the support duration of the transmitter waveforms, where $D_{\underline \varphi} \in \mathbb{N}^*$. We suppose that this duration is a finite number in order to reduce the latency and complexity.
We denote by ${E}=\mathbb{E}[|a_{mn}|^2]\ ||\underline{\varphi}||^2$, the average energy of the transmitted symbol at time $nT$ and frequency $mF$, where $\|{{ \underline \varphi}}\|^2=\sum_{q\in \mathbb{Z}}| \varphi_{q}|^2$.
\newline
Assuming a linear time-varying multipath channel $h(p,q)$, with $p$ and $q$ standing respectively for the normalized time delay and the normalized observation time, the sampled version of the received signal, $\underline r=[r_q]_{q\in\mathbb{Z}}$, has the following expression:
\begin{equation}
\underline r= \sum_{mn} a_{mn} \underline{\widetilde{\varphi}}_{mn}+ \underline n
\end{equation}
where 
\begin{equation}
\underline{\widetilde{\varphi}}_{mn}=\left[\sum_{p} h(p,q) \varphi_{mn}(q-p)\right]_{q\in \mathbb{Z}}
\end{equation}
is the channel-distorted version, $\underline{\widetilde{\varphi}}_{mn}$, of $\underline{{\varphi}}_{mn}$, and $\underline n$ is a discrete-time complex additive white Gaussian noise (AWGN), the samples of which are centered, uncorrelated with common variance $N_{0}$, where N0 is the two-sided spectral density of the original continuous-time noise. To simplify the derivation and make it tractable, while keeping the presentation general, we consider a channel with a finite number, $K$, of paths, with channel impulse response
\begin{align}
h(p,q)=\displaystyle\sum_{k=0}^{K-1}h_k e^{j 2\pi \nu_k T_{s} q}\delta(p-p_k),\nonumber
\end{align}
where $h_k$, $\nu_k$ and $p_k$ are respectively the amplitude, the Doppler frequency and the time delay of the $k^{th}$ path. The paths amplitudes $h_{k}$, $k=0..K-1$, are assumed to be centered and decorrelated random complex Gaussian variables with average powers $\pi_k=\mathbb{E}[|h_{k}|^2]$, $k=0..K-1$. In order to make the simulation tractable, the paths amplitudes $h_k$ are assumed to be i.i.d. complex Gaussian variables with zero mean and $\sum_{k=0}^{K-1}\pi_k=1$. 
\newline
The channel scattering function is therefore given by
\begin{align}
S(p,\nu)=\displaystyle\sum_{k=0}^{K-1}\pi_k \delta(p-p_k) \delta(\nu-\nu_k). \nonumber
\end{align}
At the receiver side, the decision variable on symbol $a_{kl}$ is given by
\begin{equation}
\Lambda_{kl}=\left\langle  \underline\psi_{kl}, \underline r \right\rangle =\underline\psi_{kl}^{H} \ \underline r
\label{Refdec}
\end{equation}
where $$\underline\psi_{kl}=[\psi(q-lN)e^{j2\pi\frac{kq}{Q}}]_{q\in \mathbb{Z}}$$ is the time and frequency shifted version, by $lT$ in time and $kF$ in frequency, of the reception prototype vector $\underline \psi$ used for the demodulation of $a_{kl}$. Here too, we denote by $D_{\underline\psi}$, the support duration of the receiver waveform.
\newline Here, we relax the constraint on the Tx/Rx pulses to be identical, leading to a greater flexibility in the optimization process and to an additional increase in the achievable SINR. At this point, it is worth mentioning that OFDM with CP or Zero Padding (ZP), also use different waveforms at the Tx/Rx to account for CP and ZP insertion at the transmitter respectively.
\section{POPS-OFDM Algorithm}
In this section, we will present the purpose of POPS OFDM algorithm which aims to design optimal waveforms at the Tx/Rx sides through SINR maximization, for fixed channel and synchronization imperfections statistics. Since POPS-OFDM is an iterative algorithm, it alternates between an optimization of the receiver waveform $\underline{\psi}$, for a given transmit waveform $\underline{\varphi}$, and an optimization of the transmit waveform $\underline{\varphi}$ for a given receive waveform $\underline{\psi}$.\newline Without loss of generality, we will focus on the evaluation of the signal to interference plus noise ratio (SINR) for symbol $a_{00}$. This SINR will be exactly the same for all other transmitted symbols. Referring to (\ref{Refdec}), the decision variable on $a_{00}$ can be expanded into three additive terms, as
\begin{align}
\Lambda_{00}=\underbrace{ a_{00} \left\langle  \underline\psi_{00}, \underline{\widetilde{\varphi}}_{00} \right\rangle }_{U_{00}}+\underbrace{\sum_{(m,n)\neq(0,0)} a_{mn} \left\langle \underline\psi_{00}, \underline{\widetilde{\varphi}}_{mn}\right\rangle }_{I_{00}}+ \underbrace{\left\langle  \underline\psi_{00},\underline n\right\rangle}_{N_{00}}\nonumber
\end{align}
The first term, $U_{00}$, is the useful part in the decision variable. Its power represents the useful signal power in the SINR. The second term, $I_{00}$, is the inter-symbol interference, , accounting for ISI and ICI, and the last term, $N_{00}$, is the noise term. Their respective powers represent the interference and the noise powers in the SINR.
\subsection{Average Useful Power}
The useful term in the decision variable on $a_{00}$ is given by $U_{00}= a_{00} <\underline\psi_{00}, \underline{\widetilde{\varphi}}_{00}> $. For a given realization of the channel, the average power of the useful terms is given by $P_{S}^{h}=\frac{E}{||\underline{\varphi}||^2}\  |<\underline\psi_{00}, \underline{\widetilde{\varphi}}_{00}>|^2$.
Therefore, the average of the useful power over channel realizations is $P_{S}=\mathbb{E}[P_{S}^{h}]$.
Under the notions cited above, we deduce that \begin{align}
P_S=E \ \frac{{\underline{\psi}}^{H}{\underline{\underline{KS}}}^{{{{\underline{\varphi}}}}}_{S(p,\nu)}{\underline{\psi}}}{||\underline{\varphi}||^2},\label{usefulPower}
\end{align}
where we define the useful signal Kernel matrix as \begin{align}
{\underline{\underline{KS}}}^{{{{\underline{\varphi}}}}}_{S(p,\nu)}&=\sum_{k=0}^{K-1}\pi_k \ ({{\sigma}}_{p_k}({\underline{{\varphi}}}_{00})\sigma_{p_k}({{\underline{\varphi}}}_{00})^{H})\odot  \underline{\underline{{{\Phi}}}}_{\nu_k}\nonumber \\
&=\sum_{k=0}^{K-1}\pi_k \ ({{\sigma}}_{p_k}({\underline{{\varphi}}})\sigma_{p_k}({{\underline{\varphi}}})^{H})\odot  \underline{\underline{{{\Phi}}}}_{\nu_k} 
\label{usefulkernel}
\end{align}
Given that $P_S$ is a positive entity quantity, we can state that the Kernel matrix is a positive Hermitian matrix. Since POPS-OFDM is an iterative algorithm, where $\underline{\psi}$ and $\underline{\varphi}$ have to exchange alternately their roles, it is essential to introduce this propriety relating ${\underline{\underline{KS}}}^{{{{\underline{\varphi}}}}}_{S(p,\nu)}$ and ${\underline{\underline{KS}}}^{{{{\underline{\psi}}}}}_{S(-p,-\nu)}$:
\begin{align}
&{\underline{\varphi}}^{H}{\underline{\underline{KS}}}^{{{{\underline\psi}}}}_{S(p,\nu)}{\underline{\varphi}}={\underline{\psi}}^{H}{\underline{\underline{KS}}}^{{{{\underline{\varphi}}}}}_{S(-p,-\nu)}{\underline{\psi}}.
\label{relation}
\end{align}
These equalities say that the useful signal power can be expressed as a quadratic form on $\underline{\psi}$ for a given $\underline{\varphi}$ and vice versa, given propagation channel and synchronization error statistics, summarized in the scattering function $S$.
\subsection{Average Interference Power}
The interference term within the decision variable $\Lambda_{00}$, given by ${I}_{00}=\sum_{(m,n)\ne (0,0)}a_{mn}<{\underline{\psi}}_{00},{\underline{\tilde{\varphi}}}_{mn}>$, results from the contribution of all other transmitted symbols $a_{mn}$ such that $(m,n) \ne (0,0)$. The mean power of $P_{I}^{h}$, over channel realizations, is given by
$$
P_I=\mathbb{E}[P_{I}^{h}]=\frac{E}{||\underline{\varphi}||^2} \displaystyle\sum_{(m,n)\ne (0,0)}\mathbb{E}[|<{\underline{\psi}}_{00},{\underline{\tilde{\varphi}}}_{mn}>|^2].$$
By reiterating the same derivation as the one in Section 4.1, we find that:
\begin{align}
P_{I}=E\ \frac{{\underline{\psi}}^{H}{\underline{\underline{KI}}}^{{{{\underline\varphi}}}}_{S(p,\nu)}{\underline{\psi}}}{||\underline{\varphi}||^2}.\label{interfPower}\end{align}
where the interference Kernel matrix is expressed as \begin{align}
{\underline{\underline{KI}}}^{{{{\underline\varphi}}}}_{S(p,\nu)}=\sum_{k=0}^{K-1}\pi_k \left(\sum_{(m,n)\neq (0,0)}\sigma_{p_k}({\underline{{\varphi}}}_{mn})\sigma_{p_k}({\underline{{\varphi}}}_{mn})^{H}\right) \odot \underline{\underline{{\Phi}}}_{\nu_k}.
\label{interkernel}
\end{align}
 Since $P_I$ is always positive, the interference kernel ${\underline{\underline{KI}}}^{{{{\underline{\varphi}}}}}_{S(p,\nu)}$, is also Hermitien positive semidefinite matrix. 
\newline Here too, it is important to highlight an important property that both ${\underline{\underline{KI}}}^{{{{\underline{\varphi}}}}}_{S(p,\nu)}$ and ${\underline{\underline{KI}}}^{{{{\underline\psi}}}}_{S(-p,-\nu)}$  do verify, namely
\begin{align}
&{\underline{\psi}}^{H}{\underline{\underline{KI}}}^{{{{\underline{\varphi}}}}}_{S(p,\nu)}{\underline{\psi}}= {\underline{\varphi}}^{H}{\underline{\underline{KI}}}^{{{{\underline\psi}}}}_{S(-p,-\nu)}{\underline{\varphi}}. \nonumber \\
&{\underline{\varphi}}^{H}{\underline{\underline{KI}}}^{{{{\underline\psi}}}}_{S(p,\nu)}{\underline{\varphi}}={\underline{\psi}}^{H}{\underline{\underline{KI}}}^{{{{\underline{\varphi}}}}}_{S(-p,-\nu)}{\underline{\psi}}.
\label{relation}
\end{align}
This property is very important for the following part of the paper: given any arbitrary choice of the receiver prototype vector ${\underline{\psi}}$, we can optimize the choice of the transmitter prototype vector ${\underline{\varphi}}$ through a maximization of the SINR. 
\subsection{Average Noise Power}
Foremost, we try to express the noise correlation between the two samples $N_{mn}$ and $N_{kl}$, where the noise is white:
\begin{align}
\mathbb{E}[N^{*}_{mn} N_{kl}]&=\mathbb{E}[<\underline\psi_{mn}, \underline r>^{*} \ <\underline\psi_{kl}, \underline r>]\nonumber\\
&=\underline\psi_{kl}^{H}\mathbb{E}[\underline n\  {\underline{n}}^{H}]\ \underline\psi_{mn}\nonumber\\
&=N_{0}\ \underline\psi_{kl}^{H} \ \underline\psi_{mn}\nonumber\\
&=N_{0} \ <\underline\psi_{kl},\underline\psi_{mn}>
\label{noiseEq1}
\end{align}
We remark here that the noise correlation between samples depend only on the received pulse $\underline{\psi}$ and not to the transmitted pulse $\underline{\varphi}$.
Taking $(k,l)=(m,n)$ in the previous equation, we find ..., we find that the average power of noise term, $N_{kl}$, to be equal to
\begin{align}
P_{N}&=N_{0} \left\| \underline\psi_{kl}\right\|^{2}\nonumber\\
&=N_{0} \left\| \underline\psi\right\|^{2}.
\label{noiseEq}
\end{align}
\subsection{SINR Expression}
Using the obtained expressions of the useful power $P_S$, the interference power $P_I$ and the noise power $P_N$, We can express the SINR as the following expression:
\begin{align}
&SINR=\frac{P_{S}}{P_{I}+P_{N}}=
\frac{\underline\psi^{H}\ {\underline{\underline{KS}}}^{{{{\underline{\varphi}}}}}_{S(p,\nu)} \ \underline\psi}{\underline\psi^{H} \ ({\underline{\underline{KI}}}^{{{{\underline{\varphi}}}}}_{S(p,\nu)}+\frac{{||\underline{\varphi}||^2}}{SNR} \underline{\underline{I}})\ \underline\psi},
\label{SINRFinal}
\end{align}
where $SNR=\frac{E}{N_0}$ is the Signal to Noise Ratio. This expression is valid for a general channel model without any limitations. This equation is useful in the optimization process in order to determine the received waveform, given a particular choice of the transmitted waveform at the beginning of the optimization process. \newline
We notice that, by interchanging $\underline{\varphi}$ and $\underline{\psi}$ roles, i.e. by letting $\underline{\psi}$ to be the transmitted waveform and $\underline{\varphi}$ to be the received waveform, the resulting SINR remains unchanged. In fact, using the previous identities in (\ref{SINRFinal}), we can write:
\begin{align}
&SINR=\frac{P_{S}}{P_{I}+P_{N}}=
\frac{\underline\varphi^{H}\ {\underline{\underline{KS}}}^{{{{\underline{\psi}}}}}_{S(-p,-\nu)} \ \underline\varphi}{\underline\varphi^{H} \ ({\underline{\underline{KI}}}^{{{{\underline{\psi}}}}}_{S(-p,-\nu)}+\frac{{||\underline{\varphi}||^2}}{SNR} \underline{\underline{I}})\ \underline\varphi},
\label{SINRFinal1}
\end{align}
This equation allows the design of the optimized transmitted waveform, given a particular choice of the received waveform.
Also, while maintaining the initial scattering function, $S(p,\nu)$, and by interchanging the transmitted and received waveforms and taking their time inversed versions, $\varpi{(\underline{\psi})}$,$\varpi{(\underline{\varphi})}$, we can express the SINR as:
\begin{align}
&SINR=\frac{P_{S}}{P_{I}+P_{N}}=
\frac{\varpi{(\underline{\varphi})}^{H}\ {\underline{\underline{KS}}}^{{{{\varpi{(\underline{\psi})}}}}}_{S(p,\nu)} \ \varpi{(\underline{\varphi})}}{\varpi{(\underline{\varphi})}^{H} \ ({\underline{\underline{KI}}}^{{{\varpi{({\underline{\psi}})}}}}_{S(p,\nu)}+\frac{{||\varpi(\underline{\psi})||^2}}{SNR} \underline{\underline{I}})\ \varpi{(\underline\varphi})},
\label{SINRFinal2}
\end{align}
One of the consequences of this expression is to simplify the optimization code by keeping the same code for $\underline{\psi}$ optimization given $\underline{\varphi}$, since we preserve the scattering function without any alteration. We only need to plug the time reverse of $\underline{\psi}$ in the kernel expressions to be able to obtained the corresponding time reverse of the optimum received waveform. Another consequences is that if a couple of transmitted/received waveforms ($\underline{\varphi}$, $\underline{\psi}$) achieve a given SINR, then the interchange of their roles, while taking their time reverse versions, lead to the same SINR. Last but not less important, in the maximization process, if the optimal couple $(\underline{\varphi},\underline{\psi})$, maximizing the SINR, comes to be unique, then we certainly have $\varpi{(\underline{\varphi})}=\underline{\psi}$, which means that the transmitted waveform $\underline\varphi$ and the received waveform $\underline\psi$ are reverse-time of each other. As we will see later, in this specific case, there is no way to reduce the noise correlation at the receiver by interchanging $\underline\varphi$ and $\underline\psi$.
\subsubsection{SINR expression for the conventional OFDM system}
In the previous sections, we derived the SINR expression without any type's constraints for the Tx/Rx waveforms. Hence, (\ref{SINRFinal}) is valid for whatever the transmitted and received waveforms. In particular, it is valid for the conventional OFDM system. We remind that the transmitted and received pulses for the OFDM system are expressed as follows respectively $\underline{\varphi}^{cv}=[\varphi_q^{cv}]_q$, where
\begin{align}
\varphi_q^{cv}=\left\{\begin{array}{ll}
\frac{1}{\sqrt{N}}& \mbox{ if }q=-(N-Q)\cdots 0\cdots(Q-1),\\
0 &$else$,
\end{array}
\right.
\label{phi}
\end{align}
and ${\psi}^{cv}=[\psi_q^{cv}]_q$, where
\begin{align}
\underline{\psi}^{cv}_q=\left\{\begin{array}{ll}
\frac{1}{\sqrt{Q}}& \mbox{ if }q= 0\cdots(Q-1),\\
0 &$else$.
\end{array}
\right. 
\end{align}
As is known, the received waveforms of the CP-OFDM system, when we discard the CP, constitute an orthonormal base. Similarly, it is the transmitted waveforms in the ZP-OFDM system, once we discard the ZP, constitute an orthonormal base. As a consequence, we can simply prove that the sum of the useful power (\ref{usefulPower}) and the interference power (\ref{interfPower}) is equal to:
\begin{align}
P_S^{cv}+P_I^{cv}&=\frac{E}{||\underline{\varphi}^{cv}||^2} \displaystyle\sum_{(m=0..Q-1,n)}\mathbb{E}[|<{\underline{\psi}_{00}^{cv}},{\underline{\tilde{\varphi}}_{mn}^{cv}}>|^2] \\ 
&=E\ \frac{Q}{N}
\label{somme}
\end{align}
In order to calculate the SINR ($SINR^{cv}$) for the conventional OFDM system , we need to determine the useful power, $P_S^{cv}$.
By injecting (\ref{phi}) in the useful Kernel (\ref{usefulkernel}), we obtain:
\begin{align}
\scriptstyle\underline{\underline{KS}}_{S(p,\nu)}^{\underline{\varphi}^{cv}}= \scriptstyle\left(\begin{array}{ccc}
\scriptstyle\frac{1}{N}&\cdots&\scriptstyle\sum_{k=0}^{K-1}\gamma_k(N-p_k-1)\\
\scriptstyle\sum_{k=0}^{K-1}\gamma_k(-1)&\cdots&\scriptstyle\sum_{k=0}^{K-1}\gamma_k(N-p_k-2)\\
\vdots&\ddots&\cdots\\
\scriptstyle\sum_{k=0}^{K-1}\gamma_k(-N+p_k+1)&\cdots&\scriptstyle\frac{1}{N}
\end{array}\right)
\label{KScv}
\end{align}
where
\begin{align}
\gamma_k(x)=\frac{\pi_k}{N} e^{j2\pi\nu_k T_s x}.
\label{gammaCV}
\end{align}
Hence, based on (\ref{somme}) and (\ref{KScv}), we can deduce the conventional SINR for OFDM system, which is equal to
\begin{align}
{SINR}^{cv}=\frac{\frac{P_S^{cv}}{E}}{\frac{Q}{N}-({\frac{P_S^{cv}}{E}})+\frac{1}{SNR}}
\label{SINRCV}
\end{align}
where the useful power, ${P_S^{cv}}$, using (\ref{gammaCV}), is expressed as follows:
\begin{align}
\scriptstyle P_{S}^{cv}=\scriptstyle\left\{\begin{array}{ll}
\scriptstyle E\scriptstyle\sum_{k=0}^{K-1}\gamma_k(0)+\scriptstyle E \scriptstyle\sum_{k=0}^{K-1}\scriptstyle\sum_{r=1}^{Q-1}\frac{2(Q-r)}{Q}\Re\{\gamma_k(r)\}&\\\scriptstyle \mbox{ if } \scriptstyle\displaystyle \max_{k=0..K-1}p_k\leq N-Q&\\
&\\
&\\
\scriptstyle E \scriptstyle\sum_{k=0}^{K-1}\frac{N-p_k}{Q}\gamma_k(0)
+\scriptstyle\sum_{k=0}^{K-1}\scriptstyle \left(\scriptstyle\sum_{r=1}^{N-p_k-1}\frac{2E(N-p_k-r)}{Q}\Re\{\gamma_k(r)\}\right) &\\ $else$&
\end{array}
\right. 
\label{PsC}
\end{align}

\subsubsection{Noise Correlation}
The couples ($\underline\varphi, \underline\psi$) and ($\varpi(\underline\varphi), \varpi(\underline\psi)$) provide the same SINR of the foregoing and in this sense, they are duals of each other. However, it is always the waveform at the reception that determines the noise correlation in time and frequency which is tainting the decision variables on adjacent symbols in time or/and frequency. Thus, generally, and this is also the case, CP-OFDM and its dual ZP-OFDM, the correlation provided by the couple ($\varpi(\underline\varphi), \varpi(\underline\psi)$) through $\underline\psi$ is generally different from the correlation provided by the dual ($\varpi(\underline\varphi), \varpi(\underline\psi)$) through $\varpi(\underline\varphi)$.
In light of the duality characteristic stated above, we can notice that CP-OFDM and ZP-OFDM are duals of each other. However, unlike CP-OFDM, ZP-OFDM induces induces a correlation between the noise samples in the decision variables.
\subsection{Optimization Technique}
POPS-OFDM alternates between an optimization of the transmit waveform $\underline{\varphi}$, given the receive waveform $\underline{\psi}$ and the optimization of the receive waveform $\underline{\psi}$, given the transmit waveform $\underline{\varphi}$). This is the reason why it is called the \textbf{Ping-pong Optimized Pulse Shaping} (POPS) algorithm. We remind that POPS-OFDM is an iterative algorithm. Hence, the choice of the waveform initialization is primordial and critical to be able to converge to the global maximum and find the optimal Tx/Rx waveform couple ($\underline{\psi}_{opt},\underline{\varphi}_{opt}$), that maximizes the SINR. POPS-OFDM is evaluated through Algorithm 1 for a fixed waveform initialization ($\underline{\varphi}^{(0)}$) and fixed channel parameters. 
More precisely, for the $k^{th}$ iteration, we have $\underline{\psi}^{(k-1)}$ available. We start by optimizing $\underline{\varphi}$ according to 
\begin{align}
\underline{\varphi}^{(k)}= {arg} \ \mathop{{max}}_{\underline{\varphi}}\ \ \frac{{\underline{\varphi}}^{H} \ { \underline{\underline{KS}}}^{{{{\underline{\psi}^{(k-1)}}}}}_{S(-p,-\nu)} \ {\underline\varphi}}{{\underline{\varphi}}^{H} \ {\underline{\underline{KIN}}}^{{{{\underline{\psi}^{(k-1)}}}}}_{S(-p,-\nu)}\ {\underline{\varphi}}}
\label{pb1}
\end{align}
where ${\underline{\underline{KIN}}}^{{{{\underline{\psi}}}}}_{S(-p,-\nu)}={\underline{\underline{KI}}}^{{{{\underline{\psi}}}}}_{S(-p,-\nu)}+\frac{{||\underline{\psi}||^2}}{SNR} \underline{\underline{I}}$. Then, given $\underline\varphi^{(k)}$, we carry an optimization of $\underline{\psi}$ according to 
\begin{align}
\underline{\psi}^{(k)}= {arg} \ \mathop{{max}}_{\underline{\psi}}\ \frac{{\underline{\psi}}^{H} \ {\underline{\underline{KS}}}^{{{{\underline{\varphi}^{(k)}}}}}_{S(p,\nu)} \ {\underline\psi}}{{\underline{\psi}}^{H} \ {\underline{\underline{KIN}}}^{{{{\underline{\varphi}^{(k)}}}}}_{S(p,\nu)}\ {\underline{\psi}}}
\label{pb2}
\end{align}
where ${\underline{\underline{KIN}}}^{{{{\underline{\varphi}}}}}_{S(p,\nu)}={\underline{\underline{KI}}}^{{{{\underline{\varphi}}}}}_{S(p,\nu)}+\frac{{||\underline{\varphi}||^2}}{SNR} \underline{\underline{I}}$.
\subsubsection{First Approach}
Since ${\underline{\underline{KS}}}^{{{{\underline{\varphi}}}}}_{S(p,\nu)}$ and ${\underline{\underline{KIN}}}^{{{{\underline{\varphi}}}}}_{S(p,\nu)}$ are Hermitian, symmetric, positive and semidefinite, it turns out that our problem (\ref{pb1}) amounts to a \textbf{ maximization of a generalized Rayleigh quotient} \cite{rayleigh}, which appears in many problems in engineering and pattern recognition. So, once ${\underline{\underline{KIN}}}^{{{{\underline\varphi}}}}_{S(p,\nu)}$ is invertible, our optimization problem becomes a maximization one where its solution is the eigenvector of $({\underline{\underline{KIN}}}^{{{{\underline{\varphi}}}}}_{S(p,\nu)})^{-1}\ {\underline{\underline{KS}}}^{{{{\underline{\varphi}}}}}_{S(p,\nu)}$ with maximum eigenvalue (see Algorithm \ref{algo_rrtii1}).
\begin{algorithm}[!h]
\caption{First Approach}
\label{algo_rrtii1}
\begin{algorithmic}
\REQUIRE Channel parameters ($K$, $p_k$, $\nu_k$, $T_s$, $h_k$), ${\underline{{\varphi}}}^{(0)}$, $\epsilon=10^{-10}$,
${{{\underline\psi}}}^{(0)}=(0 \cdots 0)^T$, $e^{({{{\underline\psi}}})}=e^{({{\underline\varphi}})} =10$, $k=0$, $SNR$
\STATE Compute ${\underline{\underline{KS}}}^{{{{\underline\varphi}^{(0)}}}}_{S(p,\nu)}$ and ${\underline{\underline{KI}}}^{{{{\underline\varphi}^{(0)}}}}_{S(p,\nu)}$
\WHILE{$e^{({{{\underline\psi}}})}>\epsilon$ or $e^{({{{\underline\varphi}}})}>\epsilon$}
\STATE ${\underline{\underline{KIN}}}^{{{{\underline\varphi}^{(k)}}}}_{S(p,\nu)}={\underline{\underline{KI}}}^{{{{\underline\varphi}^{(k)}}}}_{S(p,\nu)}+\frac{||\underline{\varphi}^{(k)}||^2}{SNR} \underline{\underline{I}}$
\STATE Compute ${{{\underline{\underline{\Phi}}}}}=({\underline{\underline{KIN}}}^{\underline\varphi^{(k)}}_{S(p,\nu)})^{-1}\ {\underline{\underline{KS}}}^{\underline\varphi^{(k)}}_{S(p,\nu)}$
 \STATE   Compute $[{\underline{\psi}}^{(k)},\lambda_{max}]=eig({{{\underline{\underline{\Phi}}}}})$ 
\STATE $k\gets k+1$
\STATE Evaluate ${\underline{\underline{KI}}}^{{{{\underline\psi}}}^{(k)}}_{S(-p,-\nu)}$, ${\underline{\underline{KS}}}^{{{{\underline\psi}}}^{(k)}}_{S(-p,-\nu)}$ and ${\underline{\underline{KIN}}}^{{{{\underline\psi}}}^{(k)}}_{S(-p,-\nu)}={\underline{\underline{KI}}}^{{{{{{\underline\psi}}}}}^{(k)}}_{S(-p,-\nu)}+\frac{||\underline{\psi}^{(k)}||^2}{SNR}\underline{\underline{I}}$
\STATE Compute ${{{\underline{\underline{\Theta}}}}}=({\underline{\underline{KIN}}}^{\underline\psi^{(k)}}_{S(-p,-\nu)})^{-1}\ {\underline{\underline{KS}}}^{\underline\psi^{(k)}}_{S(-p,-\nu)}$
\STATE Compute $[{{\underline\varphi}}^{(k) },\nu_{max}]=eig({{{\underline{\underline{\Theta}}}}})$
 \STATE Evaluate error: $e^{({{{\underline\psi}}})}=\|{{\underline{{\psi}}}^{(k)}}-{{\underline{{\psi}}}^{(k-1)}}\|$ and $e^{({{{\underline\varphi}}})}=\|{{\underline{{\varphi}}}^{(k)}}-{{\underline{{\varphi}}}^{(k-1)}}\|$
\ENDWHILE
\end{algorithmic}
\end{algorithm}

\subsubsection{Second approach}
Without loss of generality, through this section, we consider the optimization of the receive waveform $\underline\psi$, given the transmit waveform $\underline{\varphi}$. \newline
A possible approach to minimize the SINR consists in minimizing the denominator of (\ref{SINRFinal}), ${\underline{\psi}}^{H}{\underline{\underline{ KI}}}^{{\underline{{\varphi}}}}_{S(p,\nu)}{\underline{\psi}}$, subject to a fixed useful power, and this can be performed through the Lagrange multiplier method. This approach is useful to minimize the normalized interference power for a fixed $\frac{N_0}{E}$ value. In this case our denominator minimization problem becomes equivalent to consider the following Lagrangian function:
\begin{align}
Q^{\lambda,SNR}_{S(p,\nu)}({\underline{{\varphi}}},\underline{\psi})
&={\underline{\psi}}^{H}({\underline{\underline{  KI}}}^{{\underline{{\varphi}}}}_{S(p,\nu)}-\lambda{ \underline{\underline{ KS}}}^{{\underline{{\varphi}}}}_{S(p,\nu)}){\underline{\psi}}
\label{auxilarf}
\end{align}
where $\lambda$ is the Lagrange multiplier that depends on the Signal to Noise Ratio (SNR) value.
To assess the value around which we should choose the Lagrange multiplier, we take the gradient of the SINR with respect to $\underline{\varphi}$ or $\underline{\psi}$ .Then, we just make an identification of the obtained terms. The gradient of the SINR with respect to $\underline{\psi}$ is given by:
\begin{align}
\frac{\partial SINR}{\partial \underline{\varphi}} &=\frac{-2 {\underline{\psi}}^H {\underline{\underline{KS}}}^{{\underline{{\varphi}}}}_{S(p,\nu)}{\underline{\psi}}}{({\underline{\psi}}^{H}{\underline{\underline{ KI}}}^{{\underline{{\varphi}}}}_{S(p,\nu)}{\underline{\psi}}+\frac{N_0}{E})^2}({\underline{\underline{KI}}}^{{\underline{{\varphi}}}}_{S(p,\nu)}{\underline{\psi}} -\frac{1}{SINR}{\underline{\underline{KS}}}^{{\underline{{\varphi}}}}_{S(p,\nu)}{{\underline{\psi}}}).
\end{align}
The vector ${{\underline{\psi}}}$ leading to the optimum SINR corresponds to a null value of the gradient, i.e.
\begin{align}
{\underline{\underline{KI}}}^{{\underline{{\varphi}}}}_{S(p,\nu)}{\underline{\psi}}-\frac{1}{SINR}{\underline{\underline{KS}}}^{{\underline{{\varphi}}}}_{S(p,\nu)}{\underline{\psi}}=0.
\label{lambdaii}
\end{align}
Similarly, the vector ${{\underline{\psi}}}$ that cancels the auxiliary gradient function (\ref{auxilarf}) must verify the following equality:
\begin{align}
\frac{\partial}{\partial \underline{\psi}} Q^{\lambda,SNR}_{S(p,\nu)}({\underline{{\varphi}}},\underline{\psi})=2({\underline{\underline{ KI}}}^{{\underline{{\varphi}}}}_{S(p,\nu)}-\lambda{\underline{\underline{ KS}}}^{{\underline{{\varphi}}}}_{S(p,\nu)}){\underline{\psi}}=0.
\label{auxilsol}
\end{align}
Referring to expression (\ref{lambdaii}), the Lagrange multiplier $\lambda$ should be equal to the inverse of the SINR.
The optimization problem could be solved by the generalized eigenvalue problem (GEP), since we have to solve (\ref{auxilsol}). As our object is to maximize the SINR and as $\frac{1}{\lambda}$ is an eigenvalue of (${\underline{\underline{KI}}}^{{\underline{{\varphi}}}}_{S(p,\nu)}$,${\underline{\underline{KS}}}^{{\underline{{\varphi}}}}_{S(p,\nu)}$) in expression (\ref{lambdaii}), the optimum value $\lambda_{opt}=\frac{1}{SINR_{max}}$  corresponds to its maximum eigenvalue in order to meet our expectations. It results that ${\underline{\psi}}_{opt}$ is the eigenvector associated to the smallest eigenvalue $\lambda_{opt}$ of the GEP (${\underline{\underline{KS}}}^{{\underline{{\varphi}}}}_{S(p,\nu)}$, ${\underline{\underline{ KI}}}^{{\underline{{\varphi}}}}_{S(p,\nu)}$) (see Algorithm \ref{algo_rrtii22}).
\begin{algorithm}[!h]
\caption{Second Approach}
\label{algo_rrtii22}
\begin{algorithmic}
\REQUIRE channel parameters ($K$, $p_k$, $\nu_k$, $h_k$, $T_s$), ${\underline{{\varphi}}}^{(0)}$, $\epsilon$,
${\underline{{\psi}}}^{(0)}=0$, $e^{({\underline{{\psi}}})}=e^{(\underline{{\varphi}})} =2$, $k=0$, $SNR$
\STATE Compute ${\underline{\underline{ KS}}}^{{\underline{{\varphi}}^{(0)}}}_{S(p,\nu)}$ and ${\underline{\underline{KI}}}^{{\underline{{\varphi}}^{(0)}}}_{S(p,\nu)}$
\WHILE{$e^{({\underline{{\psi}}})}>\epsilon$ or $e^{({\underline{{\varphi}}})}>\epsilon$}
 \STATE   $\lambda=eig({\underline{\underline{ KS}}}^{{\underline{{\varphi}^{(k)}}}}_{S(p,\nu)},{\underline{\underline{KI}}}^{{\underline{{\varphi}^{(k)}}}}_{S(p,\nu)})$
 \STATE $k\gets k+1$
\STATE ${\underline{\psi}}^{(k)}$ eigenvector  associated to $\lambda_{min}$
\STATE Evaluate ${\underline{\underline{KI}}}^{{\underline{{\psi}}}^{(k)}}_{S(-p,-\nu)}$ and ${\underline{\underline{ KS}}}^{{\underline{{\psi}}}^{(k)}}_{S(-p,-\nu)}$
 \STATE   $\nu=eig({\underline{\underline{KS}}}^{{\underline{{\psi}^{(k)}}}}_{S(-p,-\nu)},{\underline{\underline{KI}}}^{{\underline{{\psi}^{(k)}}}}_{S(-p,-\nu)})$
 \STATE ${\underline{\varphi}}^{(k)}$ eigenvector  associated to $\nu_{min}$
  \STATE Evaluate errors: $e^{({\underline{{\psi}}})}=\|{{\underline{{\psi}}}^{(k+1)}}-{{\underline{{\psi}}}^{(k)}}\|$ and $e^{({\underline{{\varphi}}})}=\|{{\underline{{\varphi}}}^{(k+1)}}-{{\underline{{\varphi}}}^{(k)}}\|$
  \ENDWHILE

\end{algorithmic}
\end{algorithm}\newline
We are coming out with a Lagrange multiplier evaluation that can be characterized as a direct, simple and streamlined approach.

\subsubsection{Third Approach}
Another direct optimization method consists in diagonalizing the SINR denominator of expression (\ref{SINRFinal}) and then performing a basis change that will simplify the expression of this denominator, so that our optimization problem becomes a maximization matrix that implies finding the eigenvector of the SINR numerator that corresponds to its maximum eigenvalue. More precisely, we first introduce the Kernel function ${\underline{\underline{KIN}}}^{{\underline{{\varphi}}}}_{S(p,\nu)}={\underline{\underline{KI}}}^{{\underline{{\varphi}}}}_{S(p,\nu)}+(\frac{N_0}{E}){\underline{\underline{I}}}$. The eigen decomposition of ${\underline{\underline{KIN}}}^{{\underline{{\varphi}}}}_{S(p,\nu)}$ is ${\underline{\underline{KIN}}}^{{\underline{{\varphi}}}}_{S(p,\nu)}={\underline{\underline{U}}}\ {\underline{\underline{{{\Lambda}}}}}\ {\underline{\underline{U}}}^H$, where ${\underline{\underline{ U}}}$ is a unitary matrix, ${\underline{\underline{{{\Lambda}}}}}$ is a diagonal one with non-negative real values on the diagonal. Then, the SINR denominator can be written as ${\underline{\psi}}^{H}{\underline{\underline{ KIN}}}^{{\underline{{\varphi}}}}_{S(p,\nu)}{\underline{\psi}}={\underline{\psi}}^{H}{\underline{\underline{ U}}}\ {\underline{\underline{{{\Lambda}}}}}\ {\underline{\underline{ U}}}^H{\underline{\psi}}={\underline u}^H{\underline u}$ where ${\underline u}={\underline{\underline{{{\Lambda}}}}}^{\frac{1}{2}}{\underline{\underline{U}}}^H {\underline{\psi}}$. Since ${\underline{\underline{KI}}}^{{\underline{{\varphi}}}}_{S(p,\nu)}$ is a positive semidefinite matrix, then all the entries of ${\underline{\underline{{{\Lambda}}}}}$ are positive and greater than $\frac{N_0}{E}$which is larger than or equal to $0$. Therefore, ${\underline{\psi}}={\underline{\underline{U}}}\ {\underline{\underline{{{\Lambda}}}}}^{-\frac{1}{2}}{\underline u}$ and the SINR expression becomes
$$
SINR=\frac{{\underline u}^{H}{\underline{\underline{{{\Phi}}}}}\ {\underline u}}{{\underline u}^{H}{\underline u}},$$
where ${\underline{\underline{{{\Phi}}}}}={\underline{\underline{{{\Lambda}}}}}^{-\frac{1}{2}}{\underline{\underline{ U}}}^{H}{\underline{\underline{ KS}}}^{{\underline{{\varphi}}}}_{S(p,\nu)}{\underline{\underline{ U}}}\ {\underline{\underline{{{\Lambda}}}}}^{-\frac{1}{2}}$ is a positive matrix. Hence, maximizing the SINR is equivalent determining the maximum eigenvalue of ${\underline{\underline{{{\Phi}}}}}$ and its associated eigenvector ${\underline u}_{max}$. Hence, ${\underline{{\psi}}}^{opt}=\frac{{\underline{\underline{U}}}\ {\underline{\underline{{{\Lambda}}}}}^{-\frac{1}{2}}{\underline u}_{max}}{||{\underline{\underline{ U}}}\ {\underline{\underline{{{\Lambda}}}}}^{-\frac{1}{2}}{\underline u}_{max}||}$ (see Algorithm \ref{algo_rrtii33}).
\begin{algorithm}[!h]
\caption{Third Approach}
\label{algo_rrtii33}
\begin{algorithmic}
\REQUIRE channel parameters ($K$, $p_k$, $\nu_k$, $h_k$, $T_s$), ${\underline{{\varphi}}}^{(0)}$, $\epsilon$,
${\underline{{\psi}}}^{(0)}=0$, $e^{({\underline{{\psi}}})}=e^{(\underline{{\varphi}})} =2$, $k=0$, $SNR$
\STATE Compute ${\underline{\underline{KS}}}^{{\underline{{\varphi}}^{(0)}}}_{S(p,\nu)}$ and ${\underline{\underline{KI}}}^{{\underline{{\varphi}}^{(0)}}}_{S(p,\nu)}$
\WHILE{$e^{({\underline{{\psi}}})}>\epsilon$ or $e^{({\underline{{\varphi}}})}>\epsilon$}
\STATE ${\underline{\underline{ KIN}}}^{{\underline{{\varphi}}^{(k)}}}_{S(p,\nu)}={\underline{\underline{KI}}}^{{\underline{{\varphi}}^{(k)}}}_{S(p,\nu)}+\frac{1}{SNR} \underline{\underline{ I}}$
\STATE Compute $[\underline{\underline{{U}}},{\underline{\underline{{{\Lambda}}}}}]= eig({\underline{\underline{ KIN}}}^{{\underline{{\varphi}}^{(k)}}}_{S (p,\nu)})$
\STATE Compute ${\underline{\underline{{{\Phi}}}}}={\underline{\underline{{{\Lambda}}}}}^{-\frac{1}{2}}{\underline{\underline{ U}}}^{H}{\underline{\underline{ KS}}}^{{\underline{{\varphi}}}^{(k)}}_{S(p,\nu)}{\underline{\underline{ U}}}\ {\underline{\underline{{{\Lambda}}}}}^{-\frac{1}{2}}$
 \STATE   Compute $[{\underline u}_{max},{{\lambda}}_{max}]= eig({\underline{\underline{{{\Phi}}}}})$
\STATE $k\gets k+1$
\STATE ${\underline{\psi}}^{(k)}=\frac{{\underline{\underline{U}}}\ {\underline{\underline{{{\Lambda}}}}}^{-\frac{1}{2}}{\underline u}_{max}}{||{\underline{\underline{U}}}\ {\underline{\underline{{{\Lambda}}}}}^{-\frac{1}{2}}{\underline u}_{max}||}$
\STATE Evaluate ${\underline{\underline{KI}}}^{{\underline{{\psi}}}^{(k)}}_{S(-p,-\nu)}$, ${\underline{\underline{ KS}}}^{{\underline{{\psi}}}^{(k)}}_{S(-p,-\nu)}$ and ${\underline{\underline{ KIN}}}^{{\underline{{\psi}}}^{(k)}}_{S(-p,-\nu)}={\underline{\underline{ KI}}}^{{{{\underline{{\psi}}}}}^{(k)}}_{S(-p,-\nu)}+\frac{1}{SNR} \underline{\underline{I}}$
\STATE Compute $[{\underline V},{\underline{\underline{{{\Sigma}}}}}]= eig({\underline{\underline{ KIN}}}^{{\underline{{\psi}}}^{(k)}}_{S(-p,-\nu)})$
\STATE Compute ${\underline{\underline{{{\Theta}}}}}={\underline{\underline{{{\Sigma}}}}}^{-\frac{1}{2}}{\underline V}^{H}{\underline{\underline{ KS}}}^{{\underline{{\psi}}}^{(k)}}_{S(-p,-\nu)}{\underline V}\ {\underline{\underline{{{\Sigma}}}}}^{-\frac{1}{2}}$
\STATE Compute $[{\underline v}_{max},{{\nu_{max}}}]= eig({\underline{\underline{{{\Theta}}}}})$
 \STATE ${\underline{\varphi}}^{(k)}=\frac{{\underline V}\ {\underline{\underline{{{\Sigma}}}}}^{-\frac{1}{2}}{\underline v}_{max}}{||{\underline V}\ {\underline{\underline{{{\Sigma}}}}}^{-\frac{1}{2}}{\underline v}_{max}||}$
 \STATE Evaluate errors $e^{({\underline{{\psi}}})}=\|{{\underline{{\psi}}}^{(k+1)}}-{{\underline{{\psi}}}^{(k)}}\|$ and $e^{({\underline{{\varphi}}})}=\|{{\underline{{\varphi}}}^{(k+1)}}-{{\underline{{\varphi}}}^{(k)}}\|$
\ENDWHILE

\end{algorithmic}
\end{algorithm}\newline
It is important to note that this third approach is slower than the second one in terms of necessary compilation resources and leads approximately to the same performances but more stable numerically. However, the first approach is the fastest and the simplest approach compared to the others and leads to the same performance with more stable computation.
\section{Optimal SINR Value}
As we mentioned before, POPS-OFDM is an iterative algorithm permitting a systematic construction of the optimal waveforms at Tx/Rx sides. Unfortunately, the function to be optimized includes several local maxima in addition to one or more global maxima. As a consequence and cause of its nature, POPS-OFDM may be trapped in a local maximum, if the initialization waveform is not chosen carefully, and hence waveform initializations choice, which will be discussed in the following section, is very important. In this context, having an upper bound is beneficial to identify the waveform initialization that guarantees an optimal waveform design with high SINR.
\newline To go further in the derivation of the upper bound of the SINR, we show that we can express the exact SINR as follows:
\begin{align}
SINR=\frac{(\underline{\varphi}\otimes\underline{\psi})^H \ \underline{\underline{A}}_{S(p,\nu)} \ (\underline{\varphi}\otimes\underline{\psi})}{(\underline{\varphi}\otimes\underline{\psi})^H \ \underline{\underline{B}}_{S(p,\nu)} \ (\underline{\varphi}\otimes\underline{\psi}})
\end{align}
where
\begin{align}
 &\underline{\underline{A}}_{S(p,\nu)}=\sum_{k=0}^{K-1} \underline{\underline{\Omega}}_{k}^{(00)} \nonumber\\
&\underline{\underline{B}}_{S(p,\nu)}=\sum_{(m.n)\neq(0,0)} \sum_{k=0}^{K-1}\underline{\underline{\Omega}}_{k}^{(mn)} \nonumber\\
\end{align}
with \begin{equation}
\underline{\underline{\Omega}}_{k}^{(mn)}= \underline{\underbar{U}}_{p_{k}+nN}^{T} \ \underline{\underline{\Pi}}_{k}^{m}\  \underline{\underbar{U}}_{p_{k}+nN}, \forall (m,n) \in \mathbb{Z},
\end{equation}
with
\begin{align}
&\underline{\underline{\Pi}}_{k}^{m}=[\pi_{k} \ e^{j2\pi(\nu_{k}T_{s}+\frac{m}{Q})(q-q')}]_{q,q'\in \mathbb{Z}} \nonumber
\end{align}
\begin{align}
&{\underline{\underline{U}}}_{d}=\left[\left\{\begin{array}{ll}
1& $if$ \ q\  $mod$\  (d+m\times D\times N) = 0\\
0 & $else$
\end{array}
\right.\right]_{q,q'\in \mathbb{Z}}
\end{align}
Hence, we can conclude the upper bound of the SINR that POPS-OFDM can reach it without $\underline{\varphi}$ or $\underline{\psi}$ initializations and our problematic will be expressed as follows:
\begin{align}
SINR =\mathop{{max}}_{\underline{\chi}} \ \frac{\underline{\chi}^{H} \ \underline{\underline{A}}_{S(p,\nu)} \ \underline{\chi}}{\underline{\chi}^{H} \ \underline{\underline{B}}_{S(p,\nu)} \ \underline{\chi}}
\label{pb2}
\end{align}
where $\underline{\chi}=\underline{\varphi}\otimes\underline{\psi} = [{\varphi}_{q}\underline{\psi}]_{q \in \mathbb{Z}}$ is the Kronecker product between $\underline{\psi}$ and $\underline{\varphi}$. 
\newline 
By removing the restriction on $\underline{\chi}$ to be in the form of a Kronecker product of two vectors, $\underline{\varphi}$ and $\underline{\psi}$, and letting it to span freely the whole space, we obtain, through a maximization step, an upper bound ($\overline{SINR}$). Since $\underline{\underline{A}}_{S(p,\nu)}$ and $\underline{\underline{B}}_{S(p,\nu)}$ are symmetric, positive and semidefinite, the maximization problem in (15) turns out to be a straightforward maximization of a generalized Rayleigh quotient. Hence, the SINR upper bound is the maximum eigenvalue of $\underline{\underline{B}}_{S(p,\nu)}^{-1} \ \underline{\underline{A}}_{S(p,\nu)}$ (see Algorithm \ref{algo_rrtii}).
\begin{algorithm}[!h]
\caption{: Upper bound of the SINR}
\label{algo_rrtii}
\begin{algorithmic}
\REQUIRE Channel parameters ($K$, $p_k$, $\nu_k$, $T_s$, $h_k$), $SNR$
\STATE Compute ${\underline{\underline{A}}}_{S(p,\nu)}$ and ${ \underline{\underline{B}}}_{S(p,\nu)}$
\STATE Compute ${{{\underline{\underline{\Delta}}}}}=({ \underline{\underline{B}}}_{S(p,\nu)})^{-1}\ {\underline{\underline{A}}}_{S(p,\nu)}$
 \STATE   Compute $[{\underline{\chi}},\overline{SINR}]=eig({{{\underline{\underline{\Delta}}}}})$ 
\end{algorithmic}
\end{algorithm}
\section{Numerical Waveforms Characterization}
We consider a radio mobile channel where the scattering function $S(p,\nu)$ has a multipath power profile with an exponential truncated decaying model and classical Doppler spectrum. Let $0<b<1$ be the decaying factor, such that the paths powers can be expressed as $\pi_k=\frac{1-b}{1-b^K}b^k$. 
We recall that we work with sampled signals which also leads to use a sampling channel in time domain and therefore the Doppler spectral density, denoted by $\alpha(\nu)$, is periodic in frequency domain with period $\frac{1}{T_s}$. This scattering function obeys to the Jakes model that is decoupled from the dispersion in the time domain denoted $\beta(p)$. This means that $S(p,\nu)=\beta(p)\alpha(\nu)$, such that $\beta(p)=\sum_{k=0}^{K-1}\pi_k \delta_{K}(p-p_k)$ 
and
\begin{align}
\alpha(\nu)=\left\{\begin{array}{ll}
\frac{1}{\pi B_d}\frac{1}{\sqrt{1-(\frac{2\nu}{B_d})^2}}& \mbox{ if }|\nu|<\frac{B_d}{2}\\
0 & \mbox{if } \frac{B_d}{2} \leq |\nu|\leq\frac{1}{2T_s}
\end{array}
\right.
\label{canal}
\end{align}
where $B_d$ is the Doppler spread. 
Hence, the useful and the interference Kernel matrices, derived in (\ref{usefulkernel}) and (\ref{interkernel}), will be expressed respectively as follows:
\begin{align}
\underline{\underline{KS}}^{\underline\varphi}_{S(p,\nu)}=\sum_{k=0}^{K-1} \pi_{k} \sigma_{p_{k}}(\underline\varphi \underline\varphi^{H})\odot \underline{\underline{\Phi}}
\end{align}
\begin{align}
\underline{\underline{KI}}^{\underline\varphi}_{S(p,\nu)}=\left(\sum_n \sigma_{nN}(\sum_{k=0}^{K-1} \pi_{k} \sigma_{p_{k}}(\underline\varphi \underline\varphi^{H}))\odot \underline{\underline{\Omega}}\right)-\underline{\underline{KS}}^{\underline\varphi}_{S(p,\nu)}
\end{align}
where $\underline{\underline{\Phi}}$ and $\underline{\underline{\Omega}}$ are the Hermitian matrices for the useful and the interference kernel matrices expressed respectively as follows:
\begin{align}
\underline{\underline{{\Phi}}}&=[\int_{\nu} \alpha(\nu) e^{j2\pi\nu T_s (q-p)}]_{pq} \nonumber \\
&=[J_0(\pi B_dT_s (p-q))]_{pq}.
\end{align}
and
\begin{align}
\underline{\underline{{\Omega}}}&=[\Omega_{pq}]_{pq} .
\end{align}
with 
\begin{align}
\Omega_{pq}=\left\{\begin{array}{ll}
Q J_0(\pi B_dT_s (p-q))& $if$ \ (p-q)\  $mod$\  Q = 0\\
0 & $else$
\end{array}
\right.
\label{omega}
\end{align}
Hence, for the same context, the SINR for the conventional OFDM will be expressed as follows:
\begin{align}
{SINR}^{cv}=\frac{\frac{P_S^{cv}}{E}}{\frac{Q}{N}-(\frac{P_S^{cv}}{E})+\frac{1}{SNR}}
\label{SINRCV1}
\end{align}
where the useful power (${P_S^{cv}}$), is expressed as follows:
\begin{align}
\scriptstyle P_{S}^{cv}=\scriptstyle\left\{\begin{array}{l}
\scriptstyle\frac{E}{N}\left[1+ \scriptstyle\sum_{r=1}^{Q-1}\frac{2(Q-r)}{Q} J_0(\pi B_dT_s r)\right]\ \mbox{ if } \displaystyle \max_{k=0..K-1}p_k \leq (N-Q)\\
\\
\scriptstyle\frac{E}{N}\left[\scriptstyle \sum_{k=0}^{K-1}\frac{N-p_k}{Q}\pi_k+\scriptstyle\sum_{k=0}^{K-1} \pi_k(\scriptstyle\sum_{r=1}^{N-p_k-1}\frac{2(N-p_k-r)}{Q}J_0(\pi B_dT_s r))\right]\\ $else$
\end{array}
\right. 
\label{PsC1}
\end{align}
\subsection{POPS-OFDM Implementation Methodology}
\begin{figure}
\centering
\subfigure[Taking into account the lattice periodic structure and repetitive structure in frequency and the channel Doppler spread]{
\includegraphics [scale=0.3]{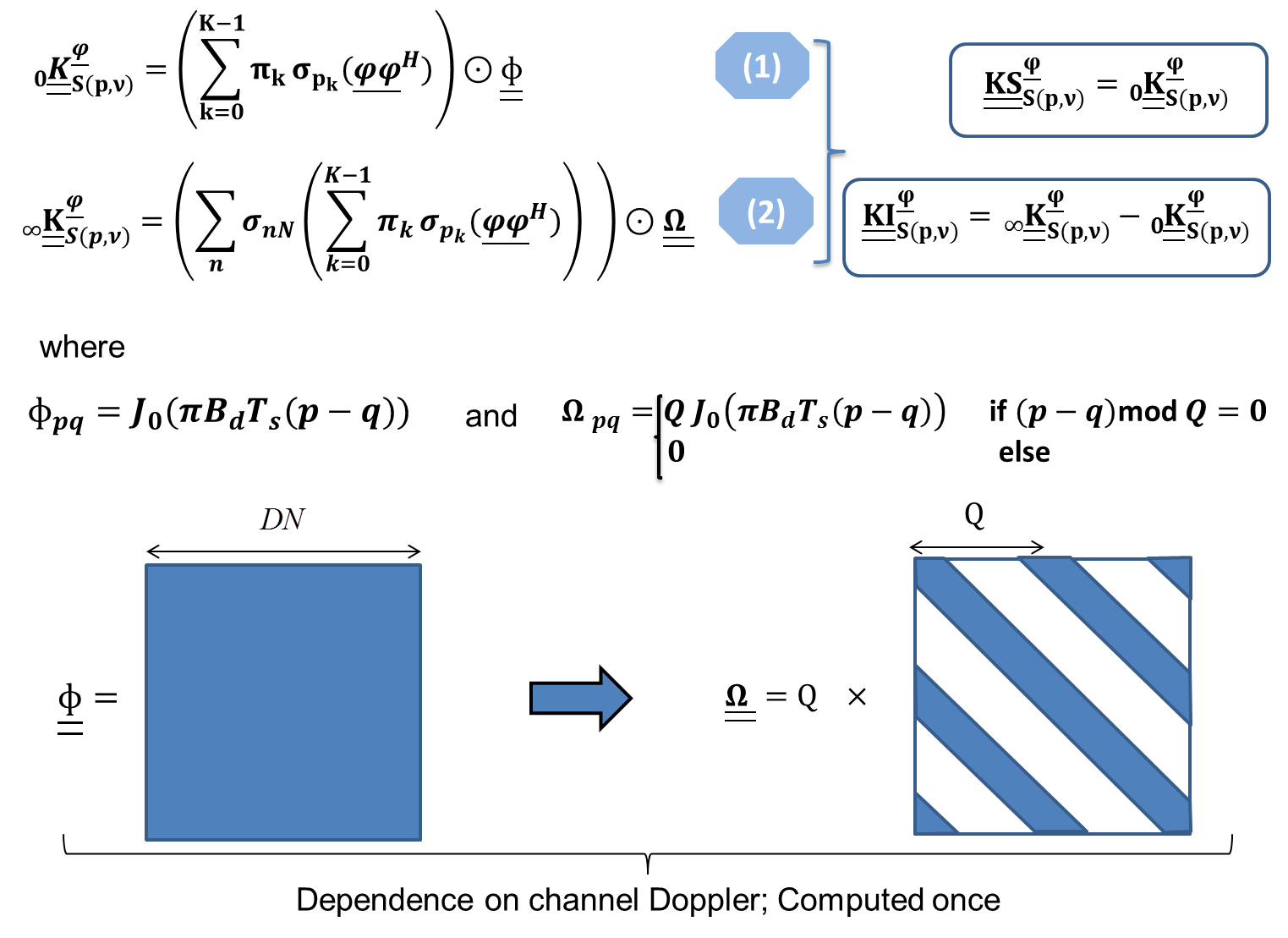}
\label{figAlgo1}
}
\hspace{5cm}
\subfigure[Taking into account the lattice periodic structure and repetitive structure in time and the channel Doppler spread]{
\includegraphics [scale=0.3]{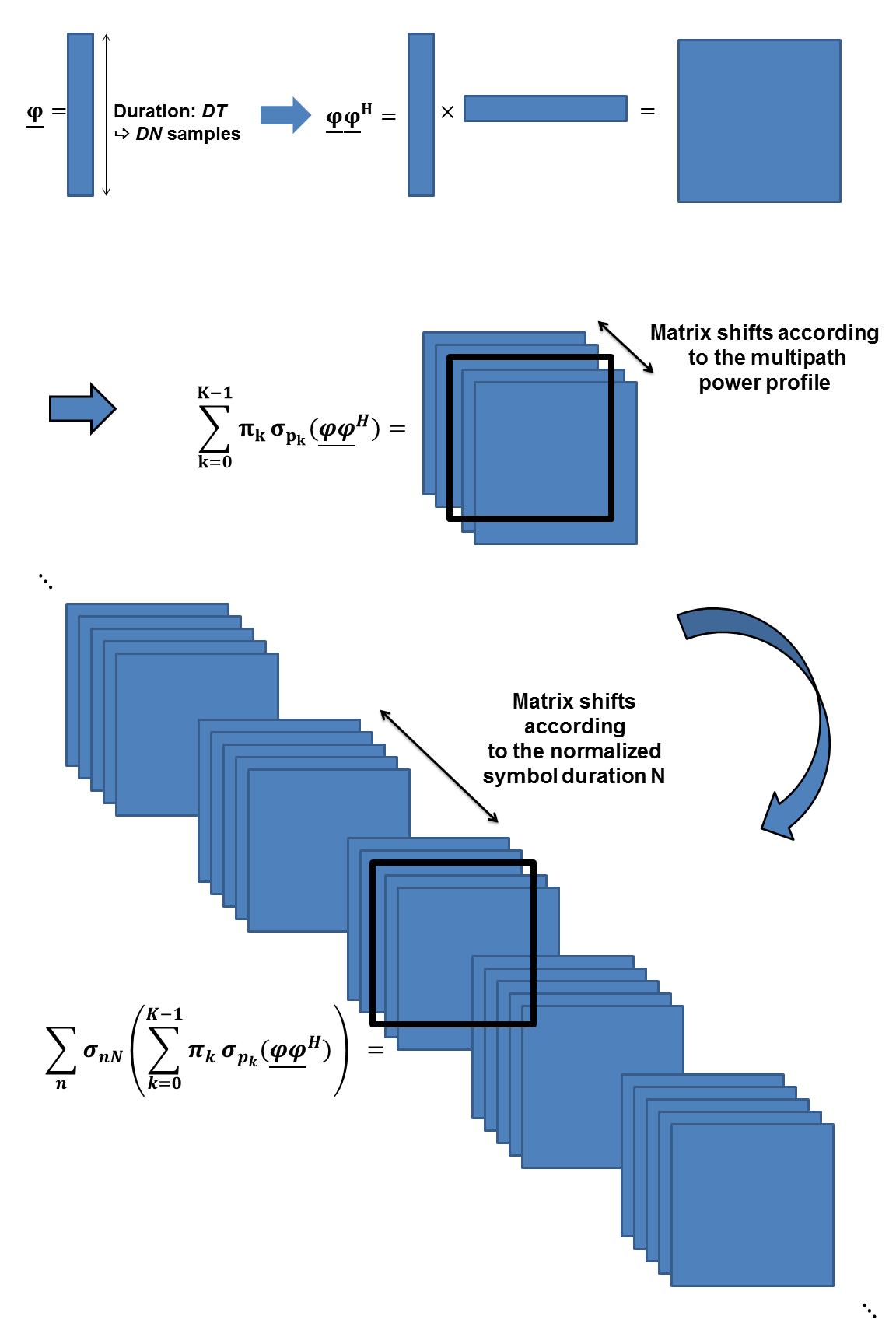}
\label{figAlgo2}}
\caption{POPS-OFDM implementation methodology}
\label{figAlgo}
\end{figure}

Through this section, we highlight a very important point which is the simplicity in the implementation of POPS-OFDM algorithm. In fact, as we can see in Fig.\ref{figAlgo1}, the matrices that depended on the Doppler channel will be computed one time for both useful and interference matrices. Then, as is depicted in Fig.\ref{figAlgo2}, we calculate the dispersion in the time according to the multipath power profile. After that, we select the matrix which has the highest energy. Hence, we can deduce the useful kernel matrix using the first formula presented in Fig.\ref{figAlgo1}. Furthermore, we shift the found matrix according to the normalized symbol duration $N$ (See Fig.\ref{figAlgo2}). Then, we select, as usual, the matrix with the highest energy to be used in the calculation of the interference kernel matrix, based on the second formula depicted in Fig.\ref{figAlgo1}.
\newpage
\subsubsection{POPS-OFDM with different Tx/Rx Pulse Shape Durations}
Many researchers shed lights on the adjacent channel interference which is caused by both transmitter non-idealities and imperfect receiver filtering \cite{ALCR1}. This type of interference need to be reduced because it contributes in network performance degradation \cite{ALCR1,ALCR2}. Mainly due to the transmitter non-linearity, the spectrum mask from transmitter will leak into adjacent channels. This interference is referred as the Out-of-Band (OOB) emissions in the frequency domain. This is a very important system parameter, since it is essential for the co-existence of parallel communications on adjacent channels whether pertaining to the same system or to different systems \cite{ALCR1}. Hence, in the literature, engineers define Adjacent Channel Leakage power Ratio (ACLR) parameter which is the ratio of the transmitted power to the power measured after a receiver filter in the adjacent RF channel \cite{ALCR1}. ACLR determines the allowed transmitted power to leak into the first or second neighboring carrier. Hence, large ACLR will guarantee a reduction of the adjacent channel interference. Furthermore, in the receiver side, we have additional interference from adjacent channels, since the receiver filter cannot be ideal \cite{ALCR2}. The adjacent Channel Selectivity (ACS) parameter is a measure of the receiver ability to receive a signal at its assigned channel frequency in the presence of a modulated signal in the adjacent channel. A poor ACS performance may lead to dropped calls in certain areas of the cells, also called ‘dead zones’ \cite{ALCR1}.\newline
Using a large waveform duration brings this waveform closer to the ideal and perfect filtering mask, since it increases the ACLR and the ACS in the Tx and Rx sides respectively. However, there is a trade-off between the reduction of the adjacent channel interference and  terminal power consumption and service delay (low latency requirements). 
We recall that POPS-OFDM offers the possibility to get flexible with different Tx/Rx pulse shape durations. In this context, it comes the idea to investigate POPS-OFDM with malleable Tx/Rx pulse shapes durations. Hence, the implementation methodology of the POPS-OFDM where the Tx/Rx pulse shape durations are different will be quiet different as we previously detailed for equal Tx/Rx pulse shapes durations. \newline
In order to make the illustration tractable, we suppose that the receiver waveform duration ($D_{\underline{\psi}}$) is greater than the transmitter waveform duration ($D_{\underline{\varphi}}$), i.e $D_{\underline{\psi}} > D_{\underline{\varphi}}$. 
\begin{enumerate}
	\item \textbf{\textit{"Ping" step}}: For the $k^{th}$ iteration, we have $\underline{\varphi}^{(k-1)}$ available. We start by optimizing $\underline{\psi}$ according to (\ref{pb2}).\newline
As is depicted in Fig.\ref{figAlgo21}, we calculate the dispersion in the time according to the multipath power profile. Then, we select the matrix used in Formula.3 in Fig.\ref{figAlgo22} to calculate the useful Kernel. The size of the selection is quiet related to the $D_{\underline{\psi}}$, since we are looking for the optimal receiver waveform ($\underline{\psi}$). After that, we shift the found matrix according to the normalized symbol duration $N$ (See Fig.\ref{figAlgo21}). Then, we select, as usual, the matrix with the highest energy to be used in the calculation of the interference Kernel matrix, based on Formula.4 depicted in Fig.\ref{figAlgo22}. Finally, we calculate the matrices which depend on the Doppler channel and which will be computed once for both useful and interference matrices in all the "Ping" steps (See Fig.\ref{figAlgo22}).
\begin{figure}
\centering
\subfigure[Taking into account the lattice periodic structure and repetitive structure in frequency and the channel Doppler spread]{
\includegraphics [height=7cm,width=9cm]{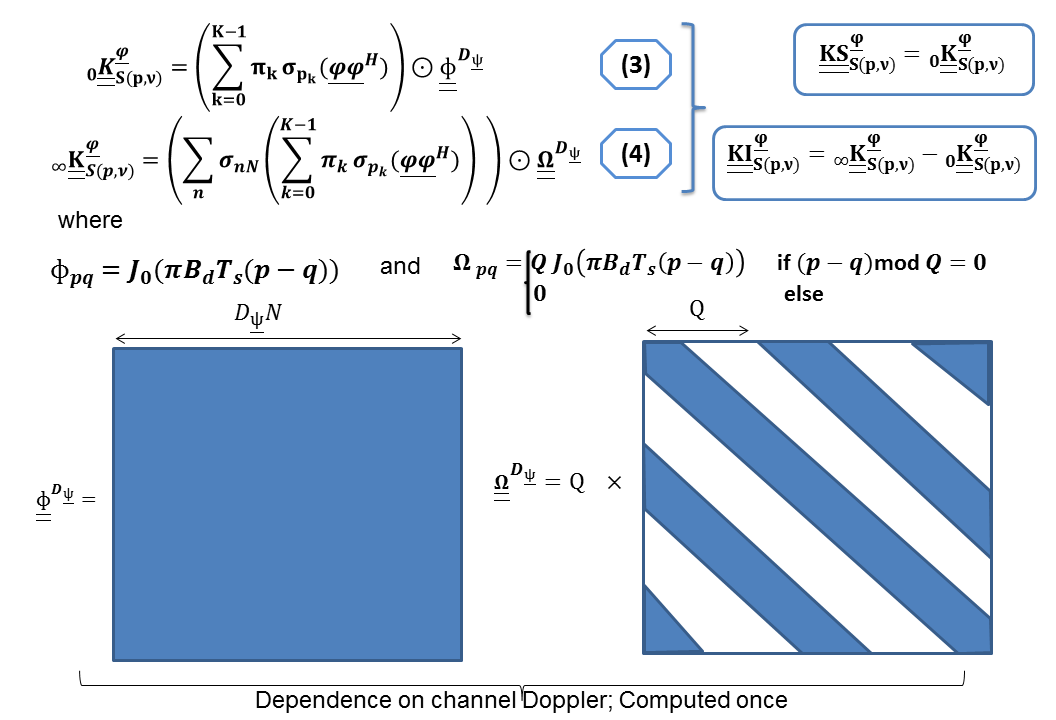}
\label{figAlgo21}
}
\hspace{5cm}
\subfigure[Taking into account the lattice periodic structure and repetitive structure in time and the channel Doppler spread]{
\includegraphics [scale=0.3]{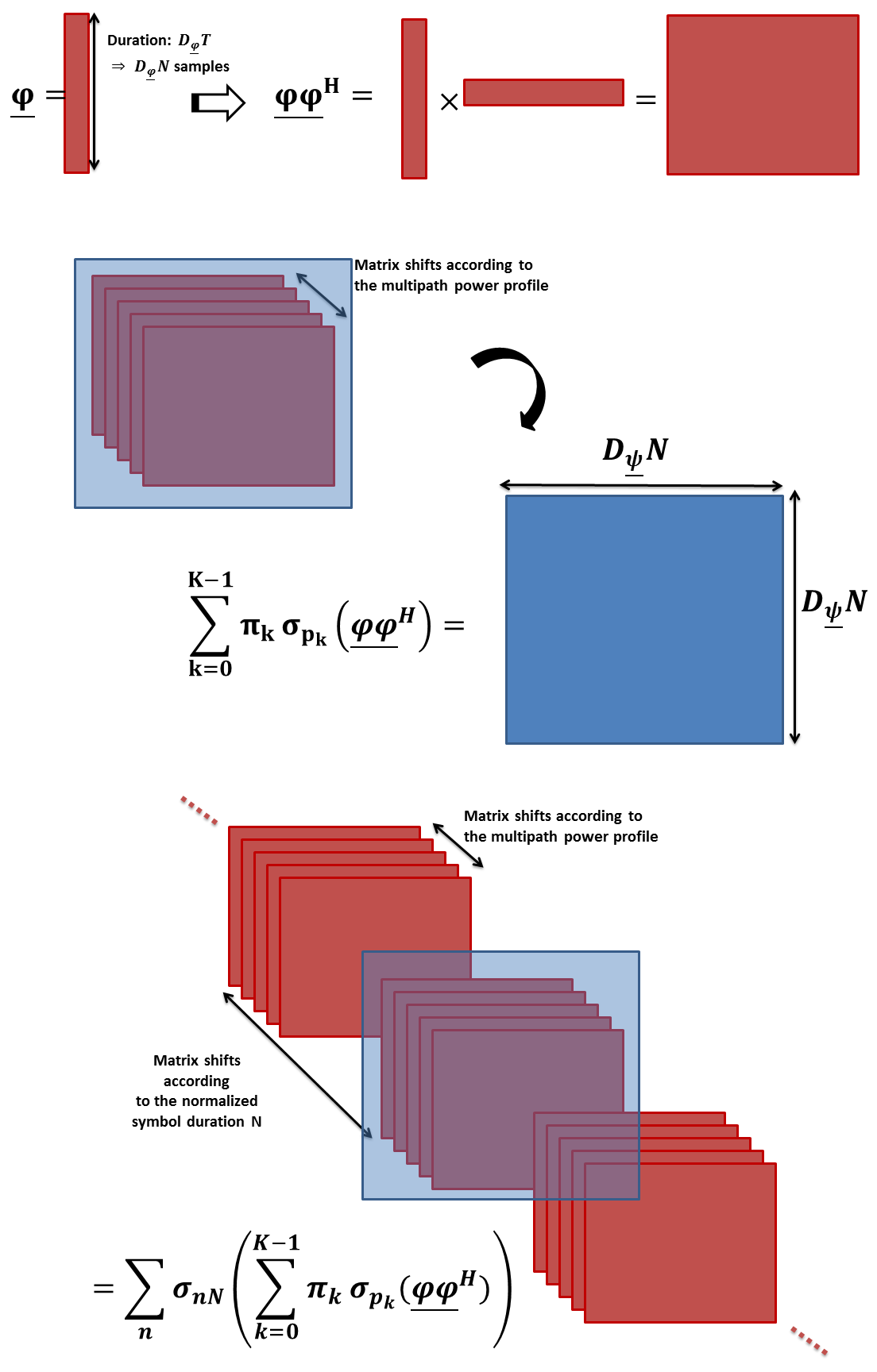}
\label{figAlgo22}}
\caption{POPS-OFDM implementation methodology ($D_{\underline{\psi}} > D_{\underline{\varphi}}$): "Ping" step}
\label{figAlgo2}
\end{figure}
\newline
\item \textit{\textbf{"Pong" step}}: For the "Pong" step, we have $\underline{\psi}^{(k)}$ available. First, we start by computing the temporel inversion of the $\underline{\psi}^{(k)}$ ($\omega(\underline{\psi}^{(k-1)})$). Then based on (\ref{SINRFinal2}), we start by optimizing $\omega(\underline{\varphi})$ according to (\ref{pb1}).
\newline The "Pong" step has the same approach as the "Ping" step when we exchange the roles between the $\underline\varphi$ and $\omega(\underline\psi)$. Hence, as is depicted in Fig.\ref{figAlgo3}, we calculate the dispersion in the time according to the multipath power profile. Then, we select the matrix used in Formula.5 in Fig.\ref{figAlgo4} in order to calculate the useful Kernel. The size of the section is quiet related to the $D_{\underline{\varphi}}$, since we are looking for the optimal receiver waveform ($\omega(\underline{\varphi})$). After that, we shift the found matrix according to the normalized symbol duration $N$ (See Fig.\ref{figAlgo3}). Then, we reiterate the same previous reasoning: We select the matrix with the highest energy to be used in the calculation of the interference kernel matrix, based on Formula.6 depicted in Fig.\ref{figAlgo4}. Finally, we calculate the matrices that depend on the Doppler channel and which will be computed one time for both useful and interference matrices in all the "Pong" steps (See Fig.\ref{figAlgo4}). Note that once we have the optimized $\omega(\underline\varphi)$, we systematically deduce $\underline\varphi$.
\begin{figure}
\centering
\subfigure[Taking into account the lattice periodic structure and repetitive structure in frequency and the channel Doppler spread]{
\includegraphics [height=7cm,width=9cm]{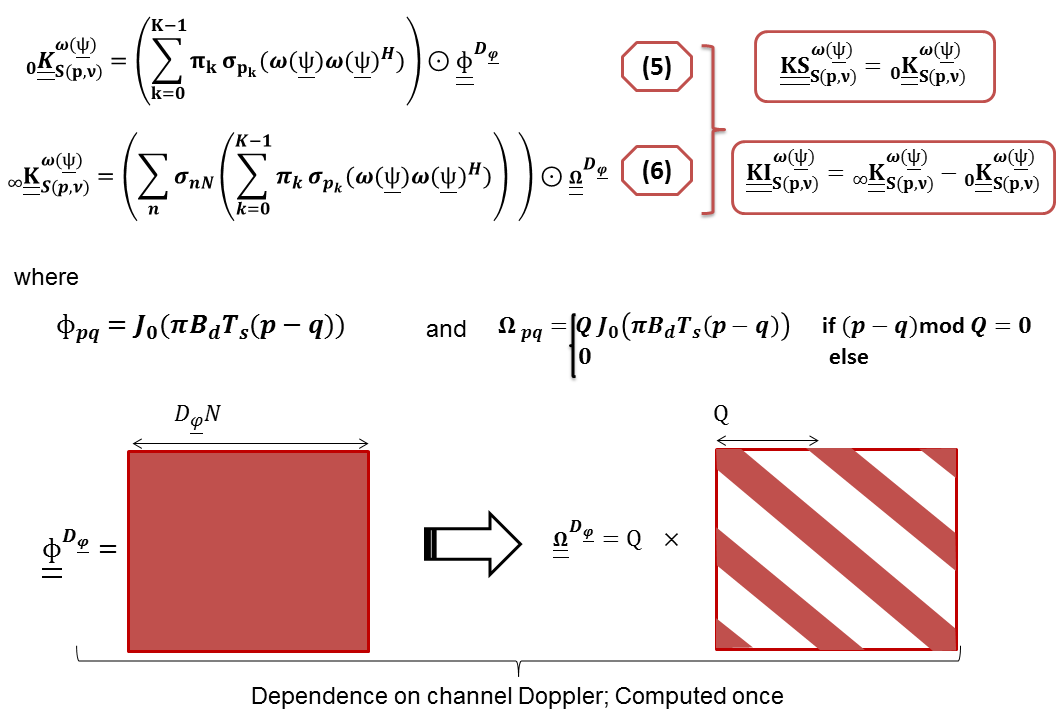}
\label{figAlgo3}
}
\hspace{5cm}
\subfigure[Taking into account the lattice periodic structure and repetitive structure in time and the channel Doppler spread]{
\includegraphics [scale=0.23]{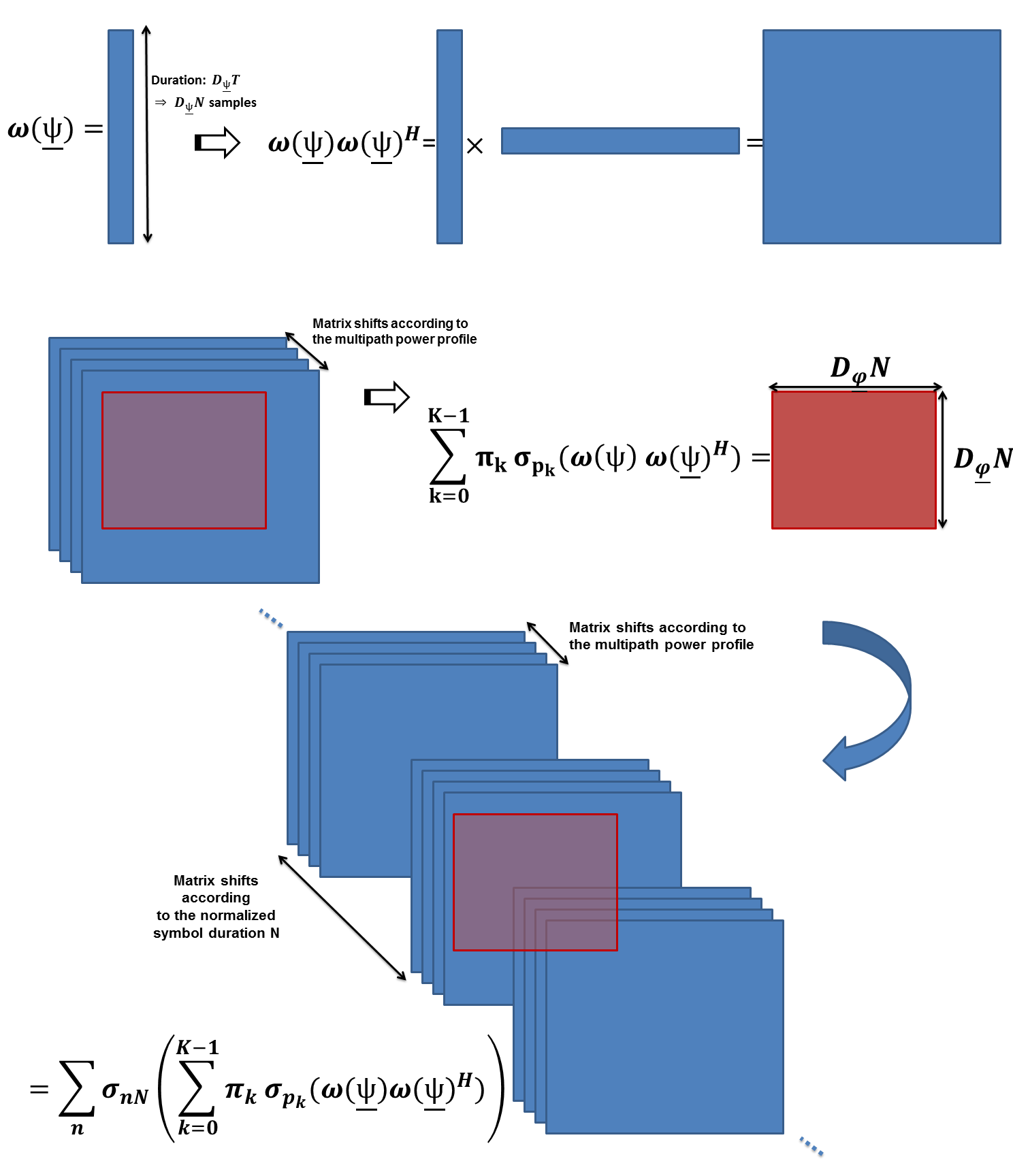}
\label{figAlgo4}}
\caption{POPS-OFDM implementation methodology ($D_{\underline{\psi}} > D_{\underline{\varphi}}$): "Pong" step}
\label{figAlgo22}
\end{figure}
\end{enumerate}
\subsection{POPS-OFDM Performance}
\begin{figure}
\centering
\subfigure[$CP=N-Q=8$.]{
\includegraphics [scale=0.4]{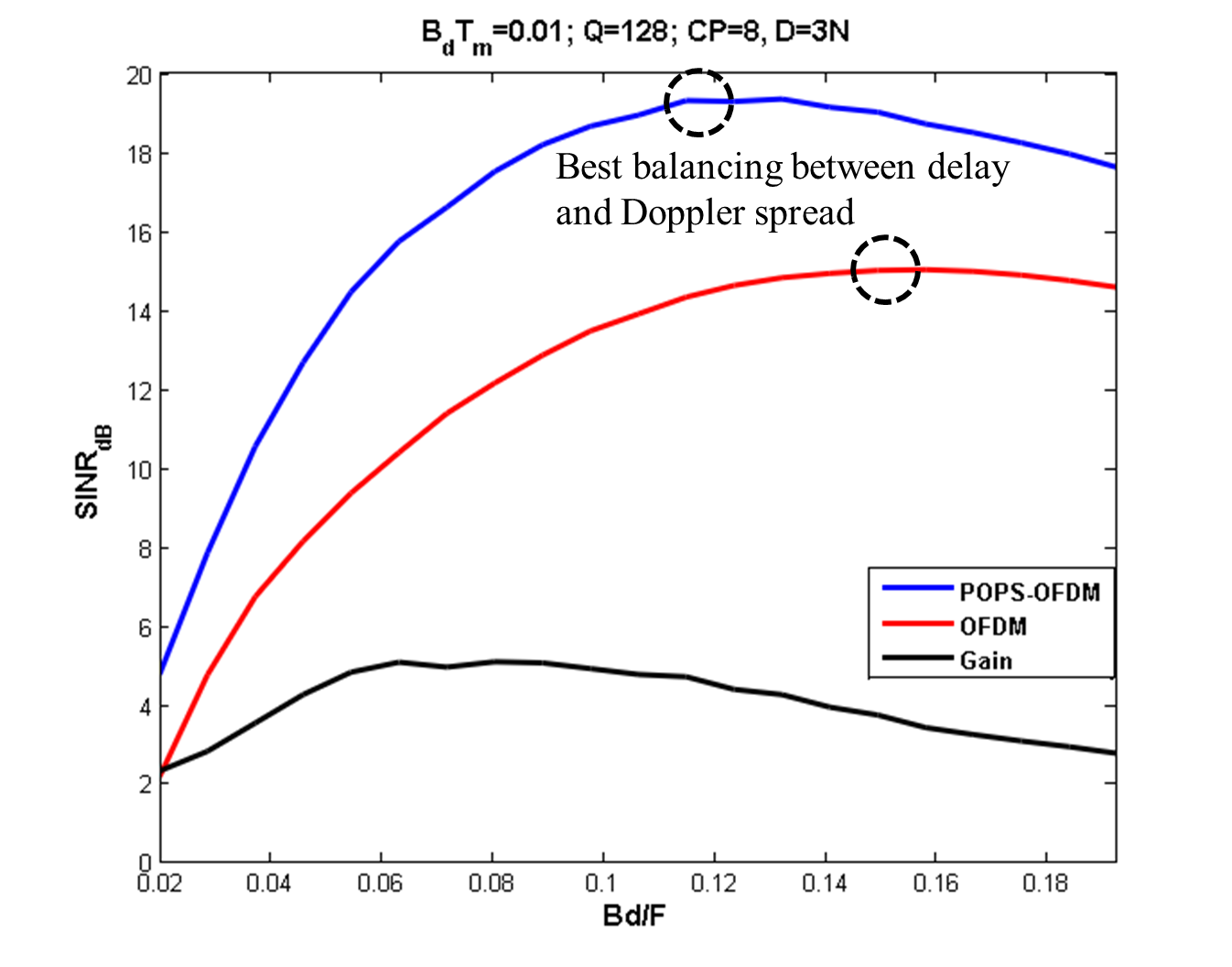}
}
\hspace{-0.3cm}
\subfigure[$CP=N-Q=32$.]{
\includegraphics [scale=0.5]{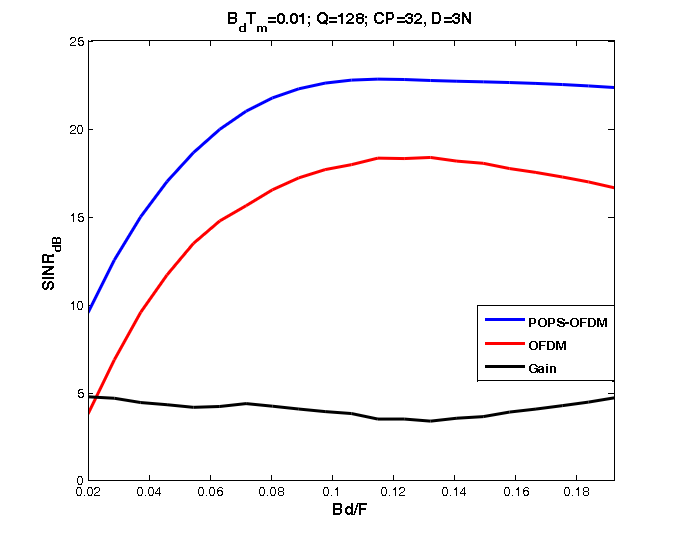}
}
\caption{ Doppler Spread-Delay Spread Balancing.
}
\label{x3}
\end{figure}

\begin{figure}
\centering
\includegraphics[height=8cm,width=10cm]{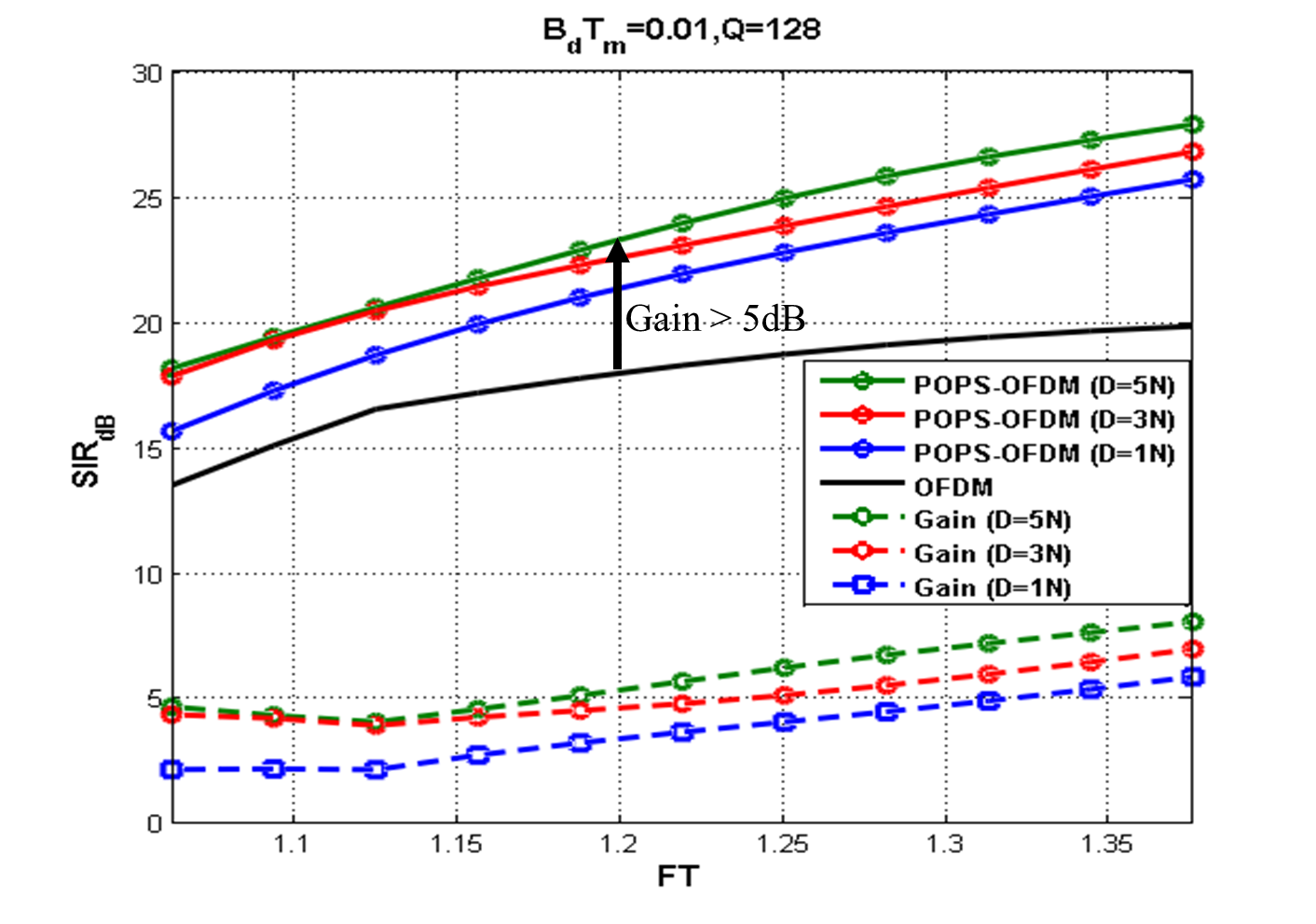}
\caption{ Performance and Gain in $SIR_{dB}$-Identical Tx/Rx Pulse Shape Durations.
}
\label{x4}
\end{figure}

\begin{figure}
\centering
\subfigure[$D=5T$.]{
\includegraphics [scale=0.5]{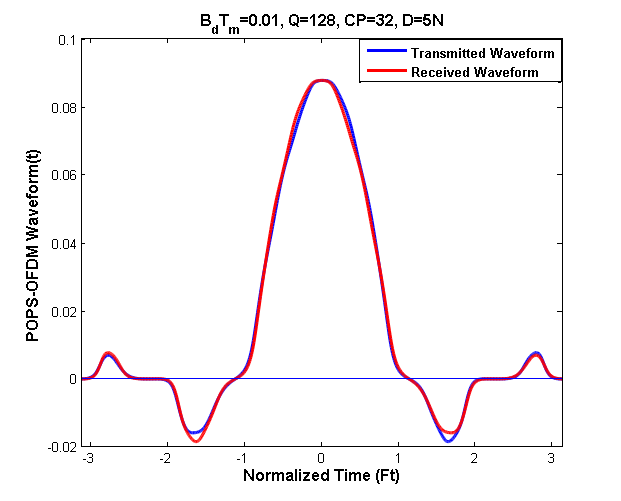}
}
\hspace{-0.3cm}
\subfigure[$D=7T$.]{
\includegraphics[scale=0.5]{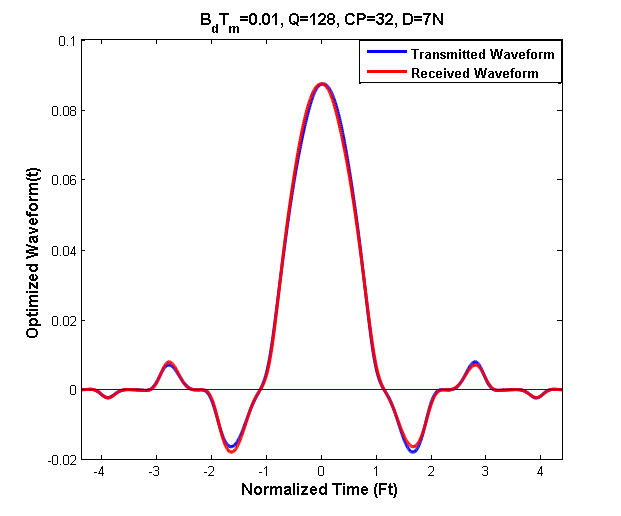}
}
\caption{ Tx/Rx Waveforms Optimization Results.
}
\label{x1}
\end{figure}

\begin{figure}
\centering
\subfigure[Spectrum of One Subcarrier.]{
\includegraphics [height=8cm,width=9cm]{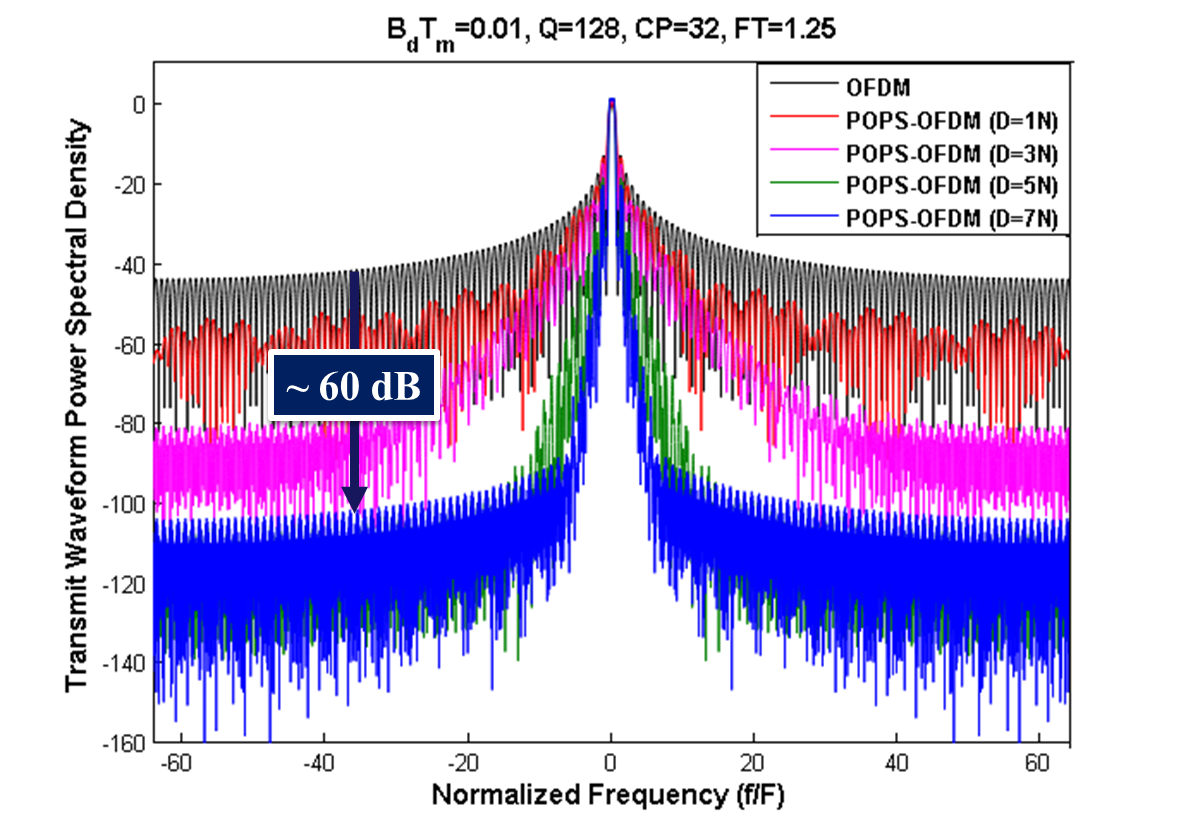}
}
\hspace{-0.3cm}
\subfigure[Spectrum of 64 Subcarriers.]{
\includegraphics[height=8cm,width=9cm]{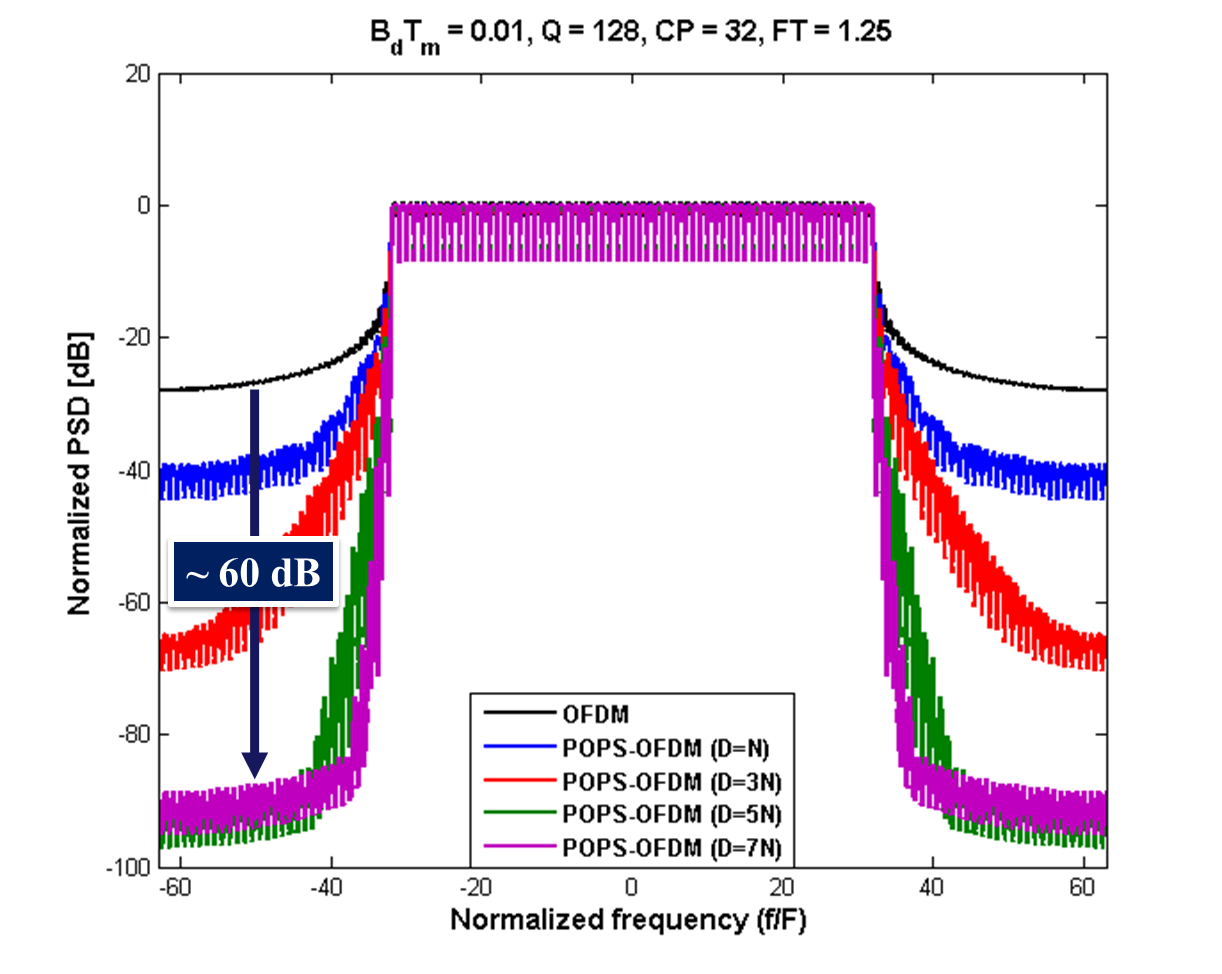}
}
\caption{ Normalized Power Spectral Density (PSD) in dB.
}
\label{x2}
\end{figure}

\begin{figure}
\centering
\subfigure[$D_\varphi=1N$.]{
\includegraphics [height=7cm,width=9cm]{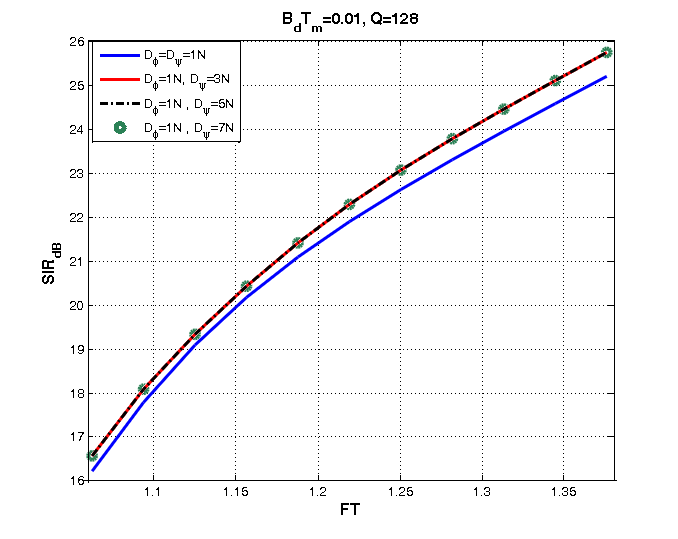}
}
\hspace{-0.3cm}
\subfigure[$D\varphi=3N$.]{
\includegraphics [height=7cm,width=9cm]{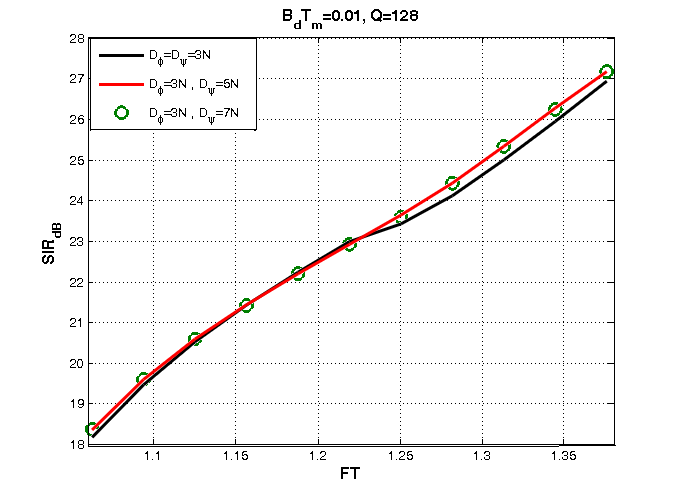}
}
\caption{ Ventilation of Complexity Between Transmitter and Receiver.}
\label{fig112}
\end{figure}

In Fig.\ref{x3}, we present the evolution of the SINR versus the normalized maximum Doppler frequency for a normalized channel delay spread values to $\frac{B_d}{F}$ where $Q=128$ and for a waveform support duration equal to $3N$. Through this simulation, we determine the Doppler spread/ delay spread balancing for a fixed channel spread value, $B_dT_m=0.01$. Also, we compare our optimized transmitter waveform design with the conventional OFDM system that deploys cyclic prefixes (CP) of $8$ or $32$ samples. This figure demonstrates that our approach outperforms the conventional OFDM system. 
Fig.\ref{x4} shows the evolution of the SIR in dB with respect to $FT$ where we compare the POPS-OFDM algorithm for different pulse shape durations with the conventional OFDM algorithm. As can be expected, the proposed system outperforms the conventional OFDM for a large range of channel dispersions, especially in the case
of a  highly  frequency  dispersive  channel and this is whatever the support duration of the waveform. This figure reveals a significant increase that can reach $8$dB in the obtained SIR when the support duration increases. Furthermore, it represents a mean to find the adequate couple $(T,F)$ of an envisaged application to insure the desired transmission quality.  Then, we can note that for a lattice density equal to $\delta=0.8$ ($FT = 1.25$), coinciding with a conventional OFDM system with a CP having one quarter of the time symbol duration, the SIR can be above $22$dB for $D=1T$.
Tx/Rx waveforms, mainly ${{{\underline\varphi}}}^{opt}$ and ${{{\underline\psi}}}^{opt}$, corresponding to the maximal SINR, are illustrated in Fig.\ref{x1} for $FT=1.25$, $B_dT_m=0.01$. This figure provides a comparison between the optimized waveforms. As we remark in this figure, the receiver and transmitter pulses are different from those of the conventional OFDM system. This confirms our claims in the sense that the conventional OFDM system not usually lead to the optimal SINR. 
Fig.\ref{x2} shows that the obtained transmitter pulse reduces exponentially of about $80$dB, the out-of-band (OOB) emissions contrary to a conventional OFDM system that requires large guard bands to do so and it can be observed that the optimal prototype waveform is more localized. We notice also that when $D$ increases the gain becomes less pronounced starting from a value of $D=5T$. More importantly, since the optimized obtained waveform reduces dramatically the spectral leakage to neighboring subcarriers, inter-user interference will be minimized especially at the uplink of OFDMA systems, where users arrive at the base station with different powers.
Fig.\ref{fig112} presents the evolution of the SIR in dB with respect to $FT$ where we compare the POPS-OFDM algorithm for different Tx/Rx pulse shape durations with the conventional OFDM algorithm for $Q=128$ and for channel spread ($B_dT_m$) equal to $0.01$.
We start by fixing the transmitter waveform duration and increase gradually the receiver waveform duration. As it is expected, when we increase the receiver waveform duration ($D_\psi$) in both cases: from $D_\psi=1N$ to $D_\psi=3N$ (See Fig.6-(a)) and from $D_\psi=3N$ to $D_\psi=5N$ (See Fig.6-(b)), the SIR is slightly superior to that obtained when we maintain the same Tx/Rx waveform durations ($D_\psi=D_\varphi=1N$ and $D_\psi=D_\varphi=3N$, respectively).
We remark also that when we consider different values of the receiver pulse shape duration which is different to the transmitter pulse shape duration, we obtain the same performance in terms of SIR. This behavior can be explained by the radical change occurred at the first increase of the receiver pulse duration. Hence, every increase after the first modification where the Tx/Rx pulse shape durations are no longer equal, the gain in terms of SIR is negligible (see Fig\ref{fig112}).
\subsection{Dependency to waveforms initializations}
\begin{figure}
\centering
\subfigure[$Q=128, CP=32, D=3T$]{
\includegraphics [width=9.5cm]{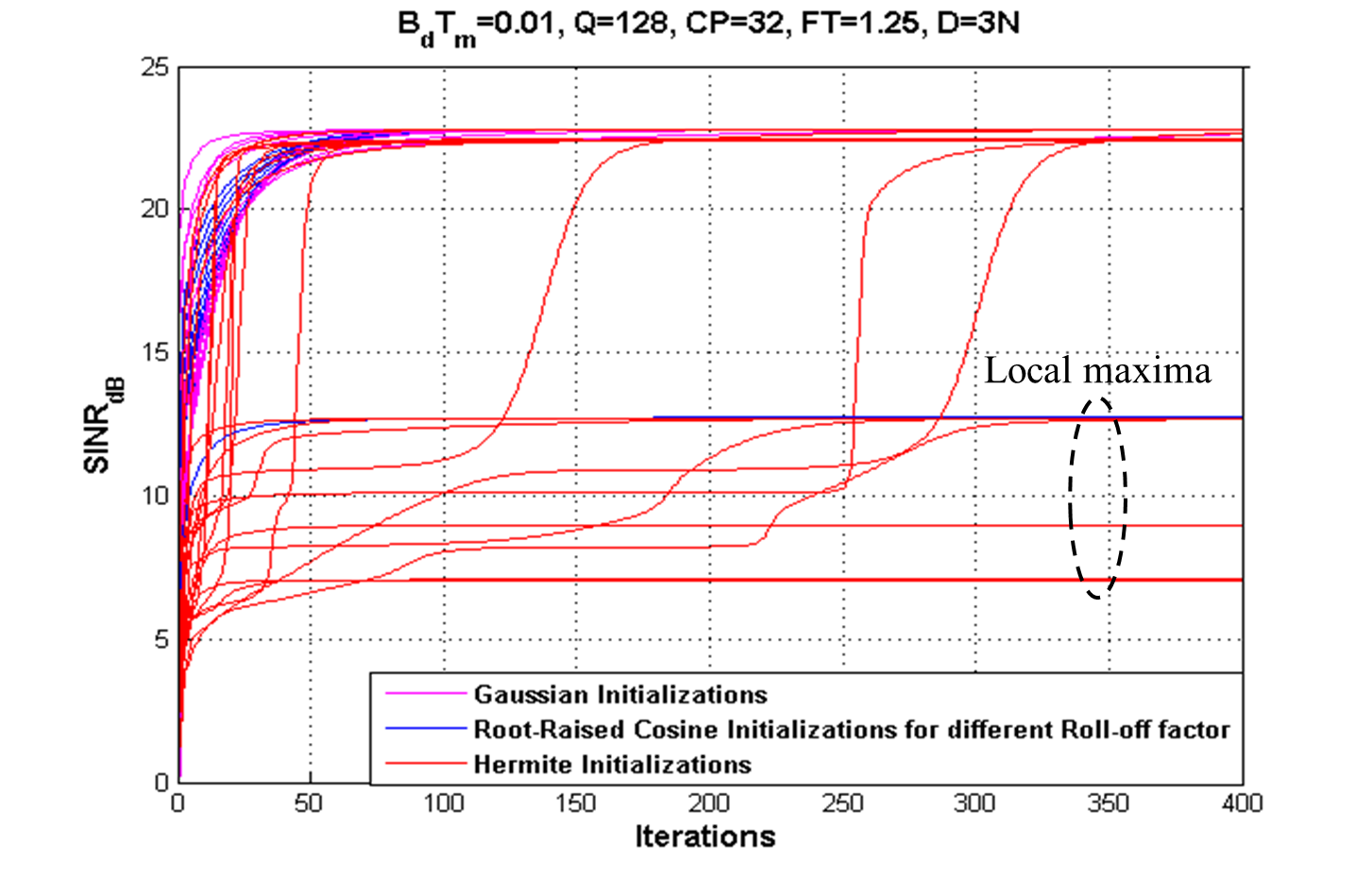}
}
\hspace{-0.3cm}
\subfigure[$Q=64, CP=16, D=1T$]{
\includegraphics [width=9.5cm]{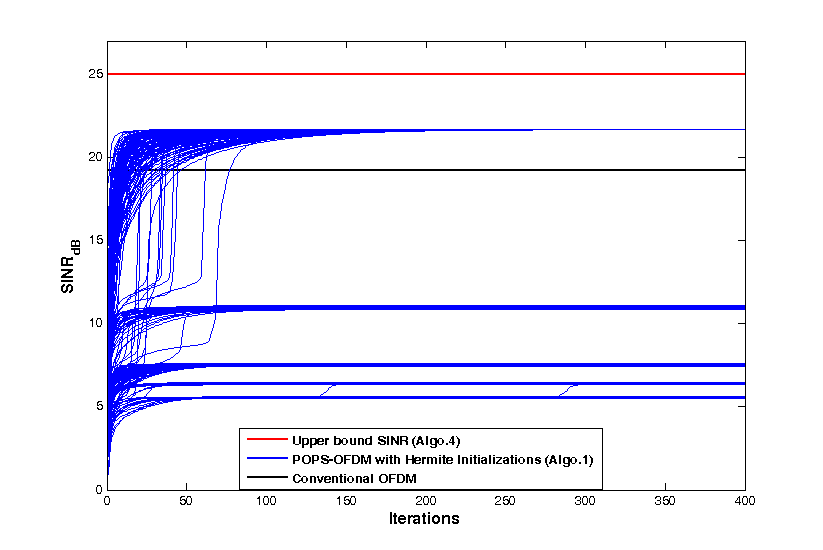}
}
\caption{  Impact of Different Waveforms Prototype Initializations.
}
\label{x6}
\end{figure}

Since POPS-OFDM banks on an iterative approach to find the optimal waveform, it is wise to study the POPS-OFDM performance in term of its sensitivity to different waveforms prototype initialization. Motivated by the fact that the Hermite functions form an orthonormal base of the Hilbert space $\mathbb L^{2}(\mathbb R)$ of square summable functions and offer in a decreasing order the best localization in time and frequency, we initiate POPS-OFDM with different linear combination of $8$ Hermite functions which are the most localized. Also, we consider gaussian waveforms where we vary the mean and the standard variation, in addition to the root-raised cosine initializations for different Roll-off factor. Fig.\ref{x6} depicts the existence of local maxima, but in the almost cases, POPS-OFDM is not trapped and converges for the same optimal maxima. Cause of hardware computation limitation, we calculate the global optimized SINR based on Algorithm 2 (see Section 5) for $Q=64$ and $D=1T$ which is drawn in Fig.$7-(b)$. Unfortunately, we conclude that whatever the considered initializations, the optimized SINR is below the global SINR target in this simulation context. But, in almost cases, it outperforms the SINR offered by the conventional OFDM system.
\subsection{Robustness characterization}
\begin{figure}
\centering
\centerline{\includegraphics[scale=0.4]{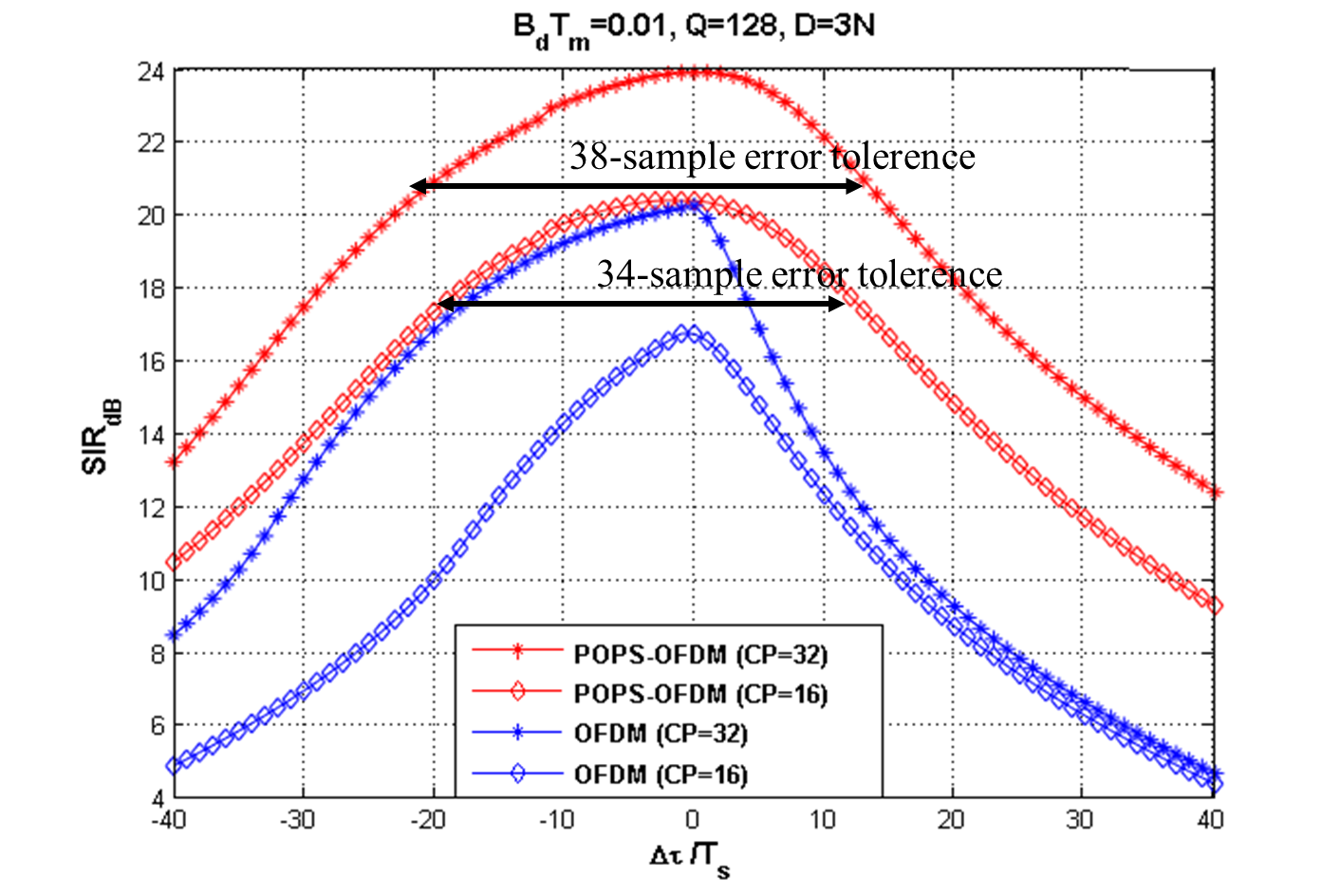}}
\caption{\label{SynchroTemps} Sensitivity to Synchronization Errors in Time.
}
\end{figure}

\begin{figure}
\centering
\centerline{\includegraphics[scale=0.5]{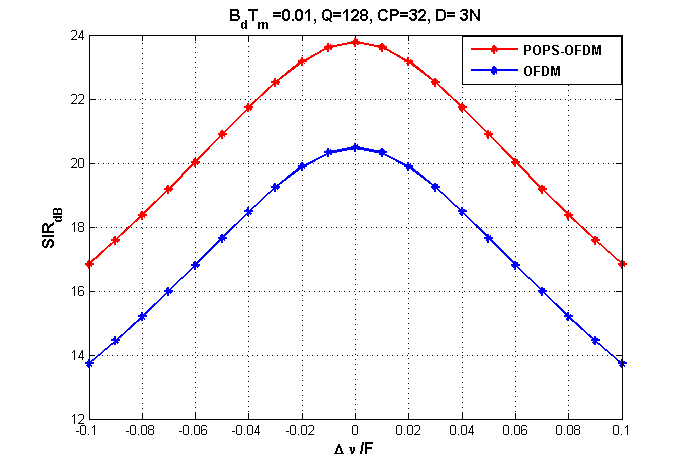}}
\caption{\label{SynchroFreq}  Sensitivity to Synchronization Errors in Frequency.
}
\end{figure}

\begin{figure}
\centering
\includegraphics[scale=0.5]{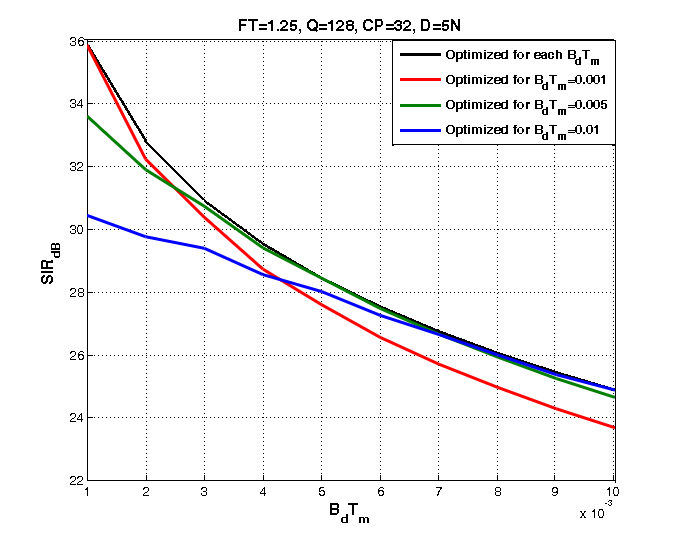}
\caption{\label{Robust3}  Sensitivity to an Estimation Error on $B_dT_m$.
}
\end{figure}

As it is known, the synchronization is a crucial indicator for efficiency of wireless communication systems and eventually for 5G \cite{ref5}, \cite{ref7}. Generally, such systems are so sensible to any synchronization error. As POPS-OFDM was principally conceived to non-orthogonal future wireless multi-carrier, it is recommended to evaluate its vulnerability against synchronization errors.  
\newline
In this section, we investigate the time and frequency synchronization errors. Then, we focus on the sensibility of the optimized waveforms for any variation around the optimal $B_dT_m$.
\newline
In Fig.\ref{SynchroTemps}, we can see clearly that the proposed algorithm outperforms the conventional OFDM in terms of robustness against the time synchronization errors when $CP=32$ and $CP=16$. For the frequency synchronization errors, the efficient proposed algorithm doesn't degrade the SIR performance compared to the conventional OFDM (See Fig.\ref{SynchroFreq}).
Fig.\ref{Robust3} illustrates the sensitivity of POPS-OFDM when we assume a synchronization error on $B_dT_m$ varying between $0.001$ and $0.01$. In this figure, we represent the SIR obtained after optimizing the waveform when $B_dT_{m_{1}}=0.001$, respectively when $B_dT_{m_{2}=0.01}$. We remark that the SINR performance of $B_dT_{m_{2}}$ degrades slowly comparing to that where the optimization is realized for $B_dT_{m_{1}}$. Therefore, it is advantageous to optimize our system for large $B_dT_m$ when we do not know its optimal value.
\section{Conclusion}
In this paper, we investigated an optimal waveform design for multicarrier transmissions over rapidly time-varying and strongly delay-spread channels. For this purpose, a novel optimizing algorithm for the transmitter and receiver waveforms is proposed. The optimized waveforms provide a neat reduction in ICI/ISI and guarantee maximal SINRs for realistic mobile radio channels. In addition to that, POPS-OFDM waveforms offer 6 orders of magnitude reduction in out-of-band emissions and reveal a great robustness to synchronization errors.
Simulation results demonstrated the excellent performance of the proposed solutions and highlighted the property of the efficient reduction of the spectral leakage obtained through the optimized waveforms.
To test the robustness of the POPS algorithm, we evaluated its sensitivity to time and frequency synchronization and also to the initialization parameters. The obtained results showed the good performance of our waveforms optimization algorithm even in high mobility propagation channels.
As such, our proposed solutions can be seen as an attractive candidate for the optimization of the spectrum allocation in 5G systems. A possible challenging research axis consists in extending the optimization for the OQAM/OFDM systems. Another interesting perspective can be investigated such that the design of OFDM pulse shapes optimized for partial equalization, for carrier aggregation and for a lower latency, with tolerant to bursty communications with relaxed synchronization.

\bibliography{IEEEabrv,ctgrefs}

\end{document}